\documentclass[MS]{dukedissertation}
\usepackage{amsmath, amssymb, amsfonts, amsthm, mathrsfs}
\usepackage{algpseudocode}
\usepackage{algorithm}
\usepackage{graphicx}
\usepackage{color}
\usepackage[usenames,dvipsnames]{xcolor}
\usepackage{bm}
\usepackage{subfigure}
\usepackage{mathabx}
\usepackage{multirow}
\usepackage{setspace}
\usepackage[english]{babel}
\usepackage[makeroom]{cancel}
\usepackage{tikz}
\usepackage{caption}
\newtheorem{theorem}{Theorem}[chapter]

\newtheorem{problem}{Problem}[chapter]
\newtheorem{lemma}{Lemma}[section]

\newtheorem{coro}{Corollary}[theorem]
\newtheorem{mydef}{Definition}[section]
\numberwithin{equation}{section}

\newtheorem{example}{Example}[section]

\newcommand\RR{\mathbb{R}}
\newcommand\ZZ{\mathbb{Z}}

\newcommand\MM{\mathbb{M}}

\newcommand\p{\partial}

\newcommand\forestdist{\text{forestdist}}
\newcommand\treedist{\text{treedist}}
\newcommand\desc{\text{desc}}

\graphicspath{{./Pictures/}}

\author{Hangjun Xu}
\title{An Algorithm for Comparing Similarity Between Two Trees}
\supervisor{Pankaj K. Agarwal}
\department{Computer Science} % Appears as Department of \department
\date{2014} % Anything but the year is ignored.

\copyrighttext{ All rights reserved except the rights granted by the\\
   \href{http://creativecommons.org/licenses/by-nc/3.0/us/}
        {Creative Commons Attribution-Noncommercial Licence}
}

\member{Kamesh Munagala}
\member{John Harer}
\usepackage[pdftex, pdfusetitle, plainpages=false, 
				letterpaper, bookmarks, bookmarksnumbered,
				colorlinks, linkcolor=black, citecolor=black,
	         filecolor=black, urlcolor=black]
				{hyperref}

\begin{document}

%-----------------------------------------------------------------------------%
% TITLE PAGE -- provides UMI abstract title page & copyright if appropriate
%-----------------------------------------------------------------------------%
\maketitle
% * <ajxuhangjun2007@gmail.com> 2015-08-13T22:33:26.932Z:
%
% 
%
%-----------------------------------------------------------------------------%
% ABSTRACT -- included file should start with '\abstract'.
%-----------------------------------------------------------------------------%
\abstract

An important problem in geometric computing is defining and computing similarity between two geometric shapes, e.g. point sets, curves and surfaces, etc. Important geometric and topological information of many shapes can be captured by defining a tree structure on them (e.g. medial axis and contour trees). Hence, it is natural to study the problem of comparing similarity between trees. We study gapped edit distance between two ordered labeled trees, first proposed by Touzet \cite{Touzet2003}. 

Given two binary trees $T_{1}$ and $T_{2}$ with $m$ and $n$ nodes. We compute the general gap edit distance in $O(m^{3}n^{2} + m^{2}n^{3})$ time. The computation of this distance in the case of arbitrary trees has shown to be NP-hard \cite{Touzet2003}. We also give an algorithm for computing the complete subtree gap edit distance, which can be applied to comparing contour trees of terrains in $\RR^{3}$.
% * <ajxuhangjun2007@gmail.com> 2015-08-13T22:33:29.070Z:
%
% 
%

%-----------------------------------------------------------------------------%
% DEDICATION -- OPTIONAL.  Put the text inside the braces.
%               (Long 'dedications' probably belong in the acknowledgements)
%-----------------------------------------------------------------------------%
\dedication{Dedicated to my parents: Sihong and Juhua.}

%-----------------------------------------------------------------------------%
% FRONTMATTER -- ToC is required, LoT and LoF are required if you have any
% tables or figures, respectively. List of Abbreviations and Symbols is 
% optional.
%-----------------------------------------------------------------------------%
\tableofcontents % Automatically generated
%\listoftables	% If you have any tables, automatically generated
\listoffigures	% If you have any figures, automatically generated
%\include{{./Abbreviations/listofabbr}} % List of Abbreviations. Start file with '\abbreviations'

%-----------------------------------------------------------------------------%
% ACKNOWLEDGEMENTS -- included file should start with '\acknowledgements'
%-----------------------------------------------------------------------------%
\acknowledgements

First and foremost, I would like to express my deepest gratitude to my advisor Professor Pankaj K. Agarwal, for suggesting this problem, and for teaching me so much about computational geometry, which have been the strongest motivation for me to pursue this master's degree en route to my Ph.D. degree in mathematics. 

I would also like to thank Professor Kamesh Munagala and Professor John Harer for being on my committee. I want to thank Marilyn Butler for helping me countlessly many times with administrative issues, and for encouraging me when I thought I couldn't complete this program. 

Last but not least, I would like to thank my Ph.D. advisor Professor Hubert Bray for being supportive, without which this project would not be possible.

%==============================================================================
%-----------------------------------------------------------------------------%
%
% MAIN BODY OF PAPER
%
%
%-----------------------------------------------------------------------------%
\chapter{Introduction}
\label{chap:introduction}

\section{Motivation}
\label{sec:motivation}

An important problem in geometric computing is \emph{shape comparison}, which concerns with \emph{defining} and \emph{computing} the similarities between two geometric shapes, e.g. point sets, curves, surfaces, etc. There are many applications of shape comparison. For instance, understanding how similar two point sets is critical in data mining and machine learning (\cite{WittenFrankHall2011}). Being able to measure the similarities between two curves help us recognize handwritings (\cite{HuangZhangKechadi2009, Su2013}) and plan motions of robots (\cite{SchwartzSharir1989, SchwartzSharir1990}). Surface matching has applications in face recognition (\cite{Delacetal2008}), image processing (\cite{LipmanDaubechies2011, Daubechies2012, Cornelisetal2011}) and even mathematical biology (\cite{Boyeretal2011}). 

The complexities of the shapes grow rapidly as their dimensions increase. To compare higher dimensional object, one technique that is often used is \emph{dimension reduction} (\cite{Fodor2002}, \cite{WittenFrankHall2011}) that ``compresses'' the objects to lower dimensions, and then perform the comparison. Many complicated shapes admit underlying tree structures that are much simpler but preserve some key topological and geometric properties of the original shapes. This suggests that we can compare these underlying tree structures and use that as a measure of similarities between the original shapes. 

Here are two examples illustrating this point. 

\subsection{Medial Axis}
\label{subsec:medial axis}

\begin{figure}[!htb]
\centering
\includegraphics[scale = .22]{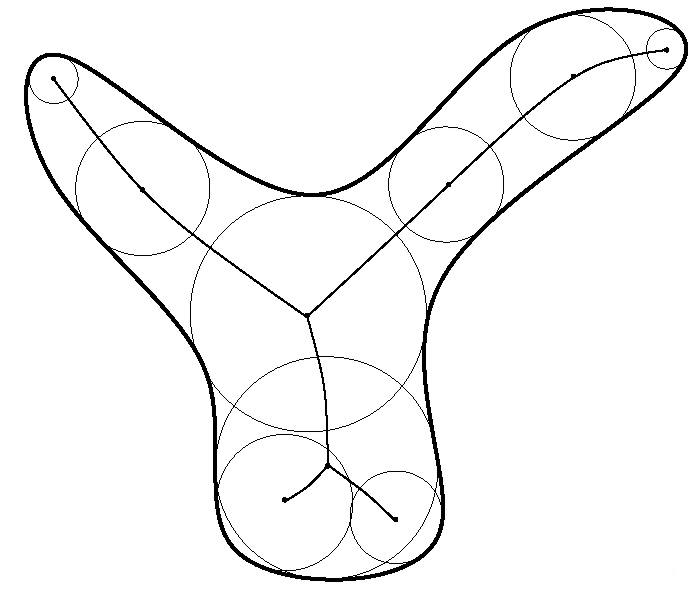}
%\vspace{-2ex}
\caption{Medial axis of a planer object. Picture from \url{http://www.lems.brown.edu/vision/Presentations/Wolter/figs.html}.}
\label{fig:medial axis}
\end{figure}

Given an object $S$ in $\RR^{n}$ with the Euclidean metric, it associated \emph{medial axis} is a set of all points $S$ that have more than one closest points on the boundary. This notion was first propose by Blum \cite{Blum1967} as a tool for shape analysis in biology. In $\RR^{2}$, suppose the boundary of $S$ is a planer curve $C$, then the medial axis consists of all centers of disks contained in $S$ that intersect $C$ tangentially at least twice (see Figure \ref{fig:medial axis}). 

In particular if $C$ is a piecewise linear polygonal curve, then the medial axis has a tree structure with vertices the same as the vertices of the boundary polygon. 

Medial axis can be viewed as a \emph{topological skeleton} of an object, which is roughly obtained by ``shrinking'' the boundary points inward until the object is deformed to a treelike object. Medial axis captures some important geometric and topological information of this object, for instance connectivity, genus, geodesics, etc. 

The medial axis of an object is often used for shape compression and shape analysis. It can also be used for shape reconstruction if both the medial axis and the radii of the disks whose centers belong to the medial axis are known (also called the medial axis transform).

In higher dimensions, we can define medial axis similarly by replacing planer disks with higher dimensional balls. Moreover, we can also use various other norms (e.g. $L^{2}$, $L^{\infty}$, etc.) which will give us different medial axises. The choice of norms depend on the particular applications. 

\subsection{Persistence and Contour Tree}
\label{subsec:persistence and contour tree}

\begin{figure}[!htb]
\centering
\includegraphics[scale = .27]{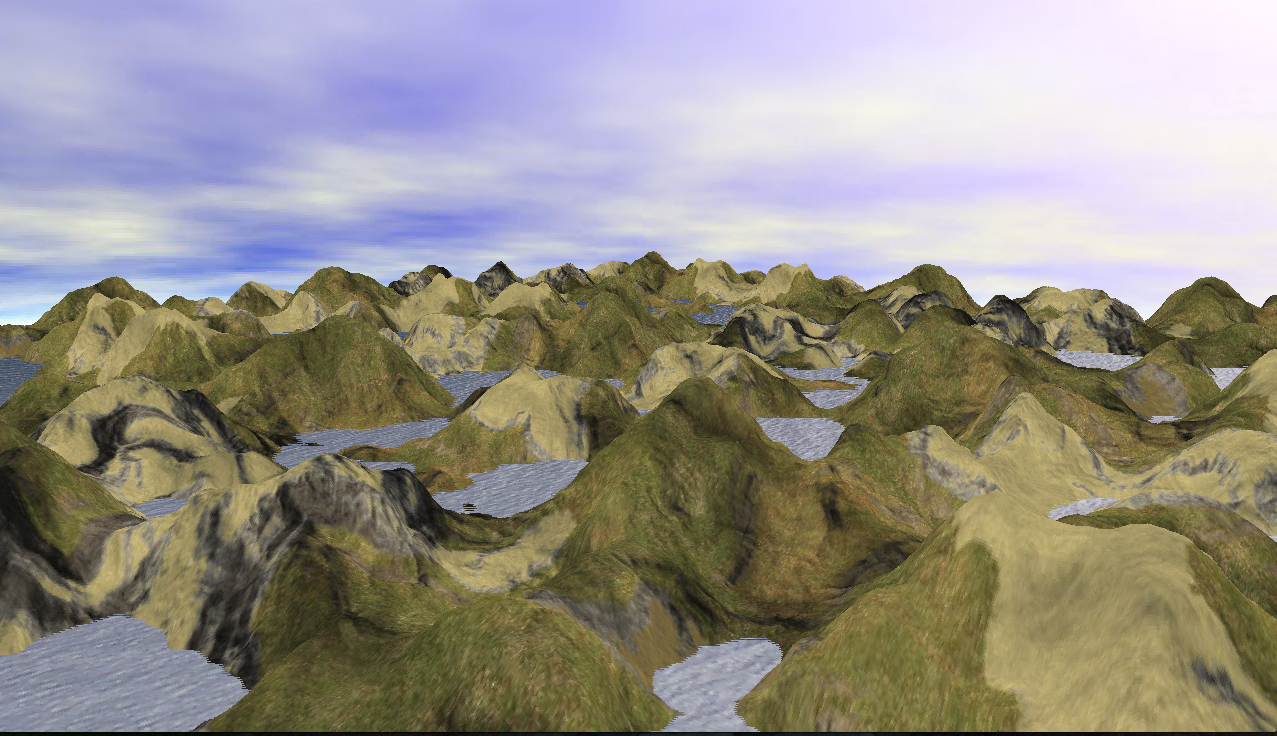}
\caption{Animated terrain drawn using polynomial height functions.}
\label{fig:animated terrain}
\end{figure}

Another example in which geometric objects have underlying tree structures is the contour tree of a terrain. A terrain in $\RR^{3}$ (see e.g. Figure \ref{fig:animated terrain}) is the graph of some function $f$ defined on $\RR^{2}$. For instance, given a triangulation $M$ of $\RR^{2}$, choose a function $f$ defined on the vertices of $M$. Linearly extend $f$ to a function on $\RR^{2}$. The graph of $f$ is then a piecewise linear triangulated surface in $\RR^{3}$. $f$ is called a height function of this terrain. Figure \ref{fig:animated terrain} is obtained by extending $f$ nonlinearly.

Now imagine that we use a plane $z = \text{constant}$ to slice this terrain, and we vary the value of $z$. The intersection of the terrain with the plane is called the \emph{level set} of height $z$. The level sets could be connected or disconnected, or empty for $z$ values that are too large or too small. In Figure \ref{fig:animated terrain} the water ponds can be viewed as the level sets (with multiple connected components) of the terrain. Notice that as we increase the $z$-values, the level sets change topology. For instance, some connected components will appear, some will disappear, and two components could merge into a single component; a single component can also split into two components. However, all these topological changes only occur at the \emph{critical points}, which are points in the domain of the height function that correspond to local maxima, local minima and saddle points. A level set first appears when our slicing plane reaches a local minimum, and ``dies'' when the plane reaches a local maximum. For saddle points that looks like the valley between two mountains, a single component would then split into two components which would disappear when the plane reaches the top of the respective mountains. 

In this slicing process, many topological information about the terrain are obtained, including its elevation data, critical point distribution, and the evolution of the level sets. The contour tree is a graph (a tree, in fact) associated with the terrain that captures these information as we slice the terrain from the bottom to the top. The nodes of the contour tree are critical points of the terrain, and there is an edge $(u, v)$ if there is a contour that appears at $v$ and disappears at $u$, where a \emph{contour} is a connected component of a level set (see Figure \ref{fig:terrain with its contour tree}).

\section{Problem Statement}
\label{sec:problem statement}

In this thesis, we compare similarity between two trees. A well-studied distance between two ordered labeled trees is the classic tree edit distance (\cite{ZhangShasha1989, ZhangShasha1997}). Edit distance measures the similarity between two trees by transforming one tree to another through \emph{pointwise} edit operations include relabeling, insertion and deletion, one node at a time. Each operation has a prescribed nonnegative cost function, and the edit distance is defined to be the minimum cost of transforming one tree to another via these operations. 

Gapped tree edit distance was first studied by Touzet \cite{Touzet2003}, in which multiple nodes, called gaps, are allowed to be inserted or deleted in a single edit operation. Moreover, the cost for such gaps is not necessarily linear. When the gap cost function is linear, gapped tree edit distance reduces to the classics edit distance. Touzet propose two models for gaps: the general model and the complete subtree model. He \cite{Touzet2003} proved that the general gap edit distance computation is NP-hard. 

The complete subtree model is rather restrictive, it is thus desirable to able to compute the general gap edit distance in certain cases. Thus, the central problem we consider is

\textbf{Problem Statement}:
Is is possible to get a polynomially computable general gap edit distance for a special class of trees, for instance, binary trees?

We answer this question in the affirmative in Chapter \ref{chap:tree edit distance with gaps}. In particular, we prove that:

\begin{theorem}[Main Theorem]
\label{thm:main theorem}

Given two ordered labeled binary trees $T_{1}$ and $T_{2}$ with vertex set $V_{1}: = V(T_{1})$ and $V_{2}: = V(T_{2})$ respectively, and an affine gap cost function. Let $m := |V_{1}|$ and $n := |V_{2}|$. The general gap edit distance between $T_{1}$ and $T_{2}$ can be computed in $O(m^{3}n^{2} + m^{2}n^{3})$ time. If $m \geq n$, then the running time is $O(m^{5})$. 
\end{theorem}

Touzet \cite{Touzet2003} gave an algorithm for computing the complete subtree gap tree edit distance. In Chapter \ref{chap:tree edit distance with gaps}, we give a different algorithm and prove that:

\begin{theorem}[Complete Subtree Gap Tree Edit Distance]
\label{thm:secondary theorem}
Given two ordered labeled binary trees $T_{1}$ and $T_{2}$ with vertex set $V_{1}: = V(T_{1})$ and $V_{2}: = V(T_{2})$ respectively, and an affine gap cost function. Let $m := |V_{1}|$ and $n := |V_{2}|$. The complete subtree gap edit distance between $T_{1}$ and $T_{2}$ can be computed in $O(m^{2}n^{2})$ time. If $m \geq n$, then the running time is $O(m^{4})$. 
\end{theorem}
\section{Outline}
\label{sec:outline and results}

In Chapter \ref{chap:a survey on tree similarities and related shape comparisons}, we study string comparison using edit distance as a motivation for tree edit distance and its generalizations. We give an overview of the classic tree edit distance, and present Zhang and Shasha's algorithm \cite{ZhangShasha1989} in detail.

In Chapter \ref{chap:tree edit distance with gaps}, we study gapped tree edit distance, and
two gap models proposed by Touzet \cite{Touzet2003}. We prove Theorem \ref{thm:main theorem} and \ref{thm:secondary theorem} using dynamic programming, which is motivated by sequence alignment algorithms and Zhang and Shasha's algorithm \cite{ZhangShasha1989}. We also discuss an application of the complete subtree gap model to terrain comparison via comparing similarity between their corresponding contour trees. 

Finally in Chapter \ref{chap:conclusion and future works}, we summarize our results, and propose some open problems suitable for future projects. 

All trees considered in this thesis are ordered and labeled.
\chapter{Classic Tree Edit Distance and Related Problems: An Overview}
\label{chap:a survey on tree similarities and related shape comparisons}

In this chapter, we study classic tree edit distance between two ordered labeled trees. 
In Section \ref{sec:introduction and motivations from sequence alignment}, we discuss the notion of edit distance and how it can be used to compare two sequences of characters. This motivates the classic tree edit distance as well as its generalizations discussed in Chapter \ref{chap:tree edit distance with gaps}. In Section \ref{sec:setups and terminologies}, we define tree edit distance and other terminologies that will be used in this and later chapters. In Section \ref{sec:zhang and shasha's algorithm} and \ref{sec:zhang and shasha algorithm complexity analysis}, we review Zhang and Shasha'a algorithm \cite{ZhangShasha1989} for computing tree edit distance in detail.

\section{Edit Distance and Sequence Alignment}
\label{sec:introduction and motivations from sequence alignment}

Edit distance was first used to measure the similarities between two strings, which are sequences of characters. It has several different definitions, and of the most commonly used variant is the \emph{Levenshtein distance}. In this version, one string is transformed to another via a sequence of edit operations that include insertion, deletion and substitution. Each edit operation has a positive cost, and the distance between two strings is defined to be the minimal cost of transforming one string to another. A graphical representation of this transformation is given by an \emph{alignment} between these two strings, which is a way of placing one string on top of another so that a one-to-one correspondence among the characters is created with deleted characters aligned with a special character denoted as a blank. 

\begin{example}\label{example:sequence alignment}
let $S_{1}$ = ``\text{save}'' and $S_{2}$ = ``\text{salvage}'', then a possible alignment is:
\begin{align}
& \text{s a - v - - e} \notag\\
&\text{s a l v a g e}
\end{align}
\end{example}

The cost of an alignment is given by the cost of the corresponding transformation. Thus, computing the edit distance is equivalent to finding the optimal alignment between two strings. 

For the cost of substituting one character with another, one can choose a metric (symmetric, positive definite and satisfies the triangle inequality) $p$ such that $p(i, j)$ is the cost of changing character $i$ to $j$. In particular, $p(i, j) = 0$ if and only if $i = j$. To penalize the deleted the characters, it is equivalent to penalize the blank characters in an alignment:

\begin{mydef}\label{def:gaps in sequence alignment}
A gap of a sequence in an alignment is a largest consecutive blank characters. 
\end{mydef}

In Example \ref{example:sequence alignment} above, we have two gaps: one of size one and another of size two. For some applications (e.g. computational biology), it is more likely to have a gap of size $k > 0$ than having $k$ isolated gaps, each of size 1 (see \cite{SetubalMeidanis1997}). Thus for the cost of gaps, it is desirable to have a function $w$ such that
\[w(k) \leq k w(1),\]
or in general 
\[w(k_{1} + k_{2}) \leq w(k_{1}) + w(k_{2}), \quad k_{1}, k_{2} \in \ZZ^{+}.\]
Such $w$ is called a \emph{convex} function. In particular:

\begin{lemma}\label{lemma:affine functions are convex}
An affine function
\begin{equation}\label{def:affine gap cost function in sequence alignment}
w(k) = 
\begin{cases}
0 & \text{for}\,\, k = 0\\
a + bk & \text{for}\,\, k \in \ZZ^{+}
\end{cases}
\end{equation}
is convex if $a \geq 0$, $b > 0$.
\end{lemma}
\begin{proof}
For any $k_{1}, k_{2}\in \ZZ^{+}$, we have
\[w(k_{1} + k_{2}) = a + b(k_{1} + k_{2}) \leq 2b +a(k_{1} + k_{2}) = w(k_{1}) + w(k_{2}),\]
since $b > 0$.
\end{proof}

Using the affine gap cost function above, if a blank character is starting a gap, we penalize it with $(a + b)$, and if it is continuing a gap, we only penalize it with $b$. Now given an alignment $M$ of strings $S_{1}$ and $S_{2}$, its cost is given by
\begin{equation}\label{def:cost of sequence alignment}
\gamma(M): = \sum_{\text{$i$ is matched with $j$, both nonblank}} p(i, j) + \sum_{g\in G}(a + b |g|),
\end{equation}
where $G$ is the set of all gaps in $M$, and $|g|$ is the size of the gap $g$.

\begin{example}
Suppose $p(i, j) = 0$ if $i = j$, and $p(i, j) = 1$ otherwise. Choose $w(k) = a + bk$. Then the alignment in example (\ref{example:sequence alignment}) has cost $(a + b) + (2a + b) = 3a + 2b$.
\end{example}

Now we focus on the problem of computing the edit distance between $S_{1}$ and $S_{2}$. Our presentation is based on \cite{SetubalMeidanis1997}. 

Let $m: = |S_{1}|, n: = |S_{2}|$, the number of characters in $S_{1}$ and $S_{2}$, respectively. Let $S_{1}[i]$ be the prefix of $S_{1}$ consists of the first $i$ characters. Define $S_{2}[j]$ similarly for $S_{2}$, $1 \leq i \leq m$, and $1 \leq j \leq n$.
To compute an optimal alignment with affine gap cost function (\ref{def:affine gap cost function in sequence alignment}), we define three auxiliary functions:
\begin{equation*}
\begin{cases}
Q_{**}[i, j]: = \text{min cost of aligning}\,\,S_{1}[i]\,\,\text{with}\,\,S_{2}[j]\,\,\text{that ends with matching}\,\,i\,\,\text{with}\,\,j  \\
Q_{\perp *}[i, j]: = \text{min cost of aligning}\,\,S_{1}[i]\,\,\text{with}\,\,S_{2}[j]\,\,\text{that ends with matching a blank}\\
 \text{node with}\,\,j \\
Q_{* \perp}[i, j] : = \text{min cost of aligning}\,\,S_{1}[i]\,\,\text{with}\,\,S_{2}[j]\,\,\text{that ends with matching}\,\,i\,\,\text{with} \\
\text{a blank node}
\end{cases}
\end{equation*}

\begin{theorem}
\label{thm:sequence alignment with gaps}
For $1 \leq i \leq m$ and $1 \leq j \leq n$, and gap cost function $w(k) = a + bk$. The matrices $Q_{**}, Q_{\perp *}$ and $Q_{*\perp}$ defined above satisfy the follow recurrence relations (with initializations given in the proof below):
\begin{equation}
Q_{**}[i, j] = p(i, j) + \min
\begin{cases}
Q_{**}[i-1, j-1]\\
Q_{\perp *}[i-1, j-1]\\
Q_{* \perp}[i-1, j-1]
\end{cases}
\end{equation}

\begin{equation}
Q_{\perp *}[i, j] = \min
\begin{cases}
Q_{**}[i, j-1] + (a+b) & \text{starting a new gap}\\
Q_{\perp *}[i, j-1] + b & \text{continuing a preexisting gap}\\
Q_{* \perp}[i, j-1] + (a+b) & \text{starting a new gap}
\end{cases}
\end{equation}

\begin{equation}
Q_{* \perp}[i, j] = \min
\begin{cases}
Q_{**}[i-1, j] + (a+b) & \text{starting a new gap}\\
Q_{\perp *}[i-1, j] + (a+b) & \text{starting a new gap}\\
Q_{* \perp}[i-1, j] + b & \text{continuing a preexisting gap}
\end{cases}
\end{equation}
The minimum cost $Q[m, n]$ of aligning $S_{1}$ with $S_{2}$ is given by
\[Q[m, n] = \min\{Q_{**}[m, n], Q_{\perp *}[m, n], Q_{* \perp}[m, n]\}.\]
\end{theorem}

\begin{proof}
We first verify the above claim for $i, j > 1$. For the recursion of $Q_{**}[i, j]$, the alignment ends with $i$ aligned with $j$, therefore no matter how $S_{1}[i-1]$ was aligned with $S_{2}[j-1]$, we simply add the penalty $p(i, j)$ to the previous cost, and there is no gap issue to worry about.

For the recursion of $Q_{\perp *}[i, j]$, the alignment ends with a blank node aligned with the node $j$. If $S_{1}[i]$ matched $S_{2}[j-1]$ ending in $i$ aligned with $j-1$, then this empty node is the beginning of a new gap, hence we penalize it with $a+b$. If a blank node was aligned with $j-1$ in the previous step, then we are continuing a preexisting gap, hence we only penalize it by $a$. The only case left is when $i$ was aligned to a blank node in the previous step, therefore this is the beginning of a new gap, hence the penalty $a+b$.

The argument for $Q_{*\perp}[i, j]$ is completely symmetric.  

Now it only left to show the above holds for $i, j = 1$. This requires us establishing appropriate initial values for $Q_{**}, Q_{\perp *}, Q_{*\perp}$:
\begin{equation}
\begin{cases}
Q_{**}[\emptyset, \emptyset] = 0\\
Q_{**}[i, \emptyset] = +\infty & \text{for}\,\, 1 \leq i \leq m\\
Q_{**}[\emptyset, j] = +\infty & \text{for}\,\, 1 \leq j \leq n
\end{cases}
\end{equation}

\begin{equation}
\begin{cases}
Q_{\perp *}[i, \emptyset] = +\infty & \text{for}\,\, 0 \leq i \leq m\\
Q_{\perp *}[\emptyset, j] = a+bj & \text{for}\,\, 1 \leq j \leq n
\end{cases}
\end{equation}

\begin{equation}
\begin{cases}
Q_{*\perp}[i, \emptyset] = a+bi & \text{for}\,\, 1 \leq i \leq n\\
Q_{*\perp}[\emptyset, j] = +\infty & \text{for}\,\, 0 \leq j \leq n
\end{cases}
\end{equation}

Here $\emptyset$ stands for a void sequence. We set $Q_{**}[i, \emptyset]$ and $Q_{**}[\emptyset, j]$ to infinity since we cannot match a nontrivial node to a node in a void sequence. Similarly $Q_{\perp *}[i, \emptyset] = \infty$ because there won't be any node in $S_{2}$ for a blank node in $S_{1}$ to be aligned to. However $Q_{\perp *}[\emptyset, j] = a+bj$ since there is a unique way to match a void sequence with a sequence with $j$ characters: use a gap with $j$ blank nodes. The initial values of $Q_{*\perp}$ are assigned in a similar manner. Since we are taking the minimum, the infinite values do not affect our computation.   
\end{proof}

Using the above recurrence relations, a straightforward dynamic programming algorithm computes the edit distance in $O(mn)$ time with $O(mn)$ space (which can be further improved to linear space, see \cite{SetubalMeidanis1997}). In the case that the gap cost function is arbitrary, similar recurrences can be obtained that compute the distance in $O(m^{2}n + mn^{2})$ time. 

A string can be thought of as a tree with a single leaf. In fact, many tree algorithms specialize to strings, and are as efficient as the best string algorithms \cite{ZhangShasha1997}. Now the question is: Can we generalize the string edit distance for strings to trees? The idea is again to transform one tree to another via a sequence of edit operations, and the distance is defined to be the minimal cost of such transformations (Precise definition see Section \ref{sec:setups and terminologies} below). In 1977, Selkow \cite{Selkow1977} first attempted to generalize string edit distance to ordered trees. Later in 1979, Tai \cite{Tai1979} gave the first definition of edit distance between two ordered labeled trees, and the first polynomial algorithm to compute it. Many variants have been extensively studied, e.g. edit distance between unordered trees \cite{ZhangStatmanShasha1992}, tree alignment problem \cite{JiangWangzhang1995}, tree inclusion problem \cite{KilpelainenMannila1995}, etc. 

Given two ordered labeld trees $T_{1}$ and $T_{2}$ with $m$ and $n$ nodes, respectively. A straightforward dynamic programming algorithm computes the edit distance in time $O(m^{2}n^{2})$. In 1989, Zhang and Shasha \cite{ZhangShasha1989} computed the edit distance in $O(mn\cdot \min\{D_{1}, L_{1}\}  \cdot  \min\{D_{2}, L_{2}\})$ time with space $O(mn)$, where $D_{i}$ is the depth of $T_{i}$, and $L_{i}$ is the number of leaves of $T_{i}$, $i = 1, 2$. However, the worst case running time is still $O(m^{2}n^{2})$. In 1998, Klein \cite{Klein1998} modified Zhang and Shasha's algorithm using path decompositions, and improved the running time to $O(m^{2}n\log n)$. In 2001, Chen \cite{Chen2001} gave an algorithm that compute the distance in $O(mn + L_{1}^{2}n + L_{1}^{2.5}L_{2})$ time. In 2003, Dulucq and Touzet \cite{DulucqTouzet2003} computed the distance in $O(mn\log^{2}n)$ time. In 2009, Demaine et al computed the edit distance in $O(m^{2}n)$ time. 

For the remainder of this chapter, we study Zhang and Shasha's algorithm in detail. 

\section{Zhang and Shasha's Algorithm Part I: Setup}
\label{sec:setups and terminologies}
In this section, we define the edit distance between two \emph{ordered labeled} rooted trees. A tree is said to be ordered if for each node, we can put a left-to-right order on its siblings. Every tree embedded in $\RR^{n}$ has a natural order after we fix an arbitrary vector and simply determine the order of the nodes with a sweep out along the direction of that vector. A tree is called labeled if each node has an assigned symbol taken from a finite alphabet $\Sigma$. 

The edit operations are
\begin{enumerate}
\item[(1)] Rename. To rename one node label to another.
\item[(2)] Delete. To delete a node $u$, and all children of $u$ become children of the parent of $u$, while maintaining the order.
\item[(3)] Insert. To insert a node $u$ as a child of $u'$. A consecutive sequence of children of $u'$ now becomes the children of $u$. 
\end{enumerate}
Represent an edit operation as a pair of nodes $(a, b)$, or as $a\rightarrow b$, to indicate that we relabel the node with $a$ by $b$. Introduce a special label $\lambda$ that is not in $\Sigma$, so that $(a, \lambda)$ or $a\rightarrow\lambda$ indicates the deletion of $a$, and similarly $(\lambda, b)$ or $\lambda\rightarrow b$ indicates the insertion of $b$. We sometimes identify a tree node with its label whenever there is no ambiguity. Consider two trees $T_1$ and $T_2$. Let $V(T_1)$ and $V(T_2)$ be the respective vertex set. Given a distance function (positive definite, symmetric, and satisfies the triangle inequality) $\gamma: V(T_1)\cup \{\lambda\} \times V(T_2)\cup\{\lambda\} \longrightarrow \RR^{\geq 0}$, we call such $\gamma$ a \emph{cost function} on edit operations. Now let $S$ be a sequence $s_{1}, s_{2}, \cdots, s_{k}$ of edit operations, and each of them has a cost, and thus the cost of $S$ is the sum of the all the costs:
\begin{equation}
\gamma(S): = \sum_{i = 1}^{|S|} \gamma (s_i),
\end{equation}
where $|S|$ denotes the number of edit operations in $S$. The tree edit distance is defined as:
\begin{mydef}\label{def:tree_edit_distance}
$\delta(T_1, T_2):= \min\{\gamma(S) | S\,\, \text{is an edit operation sequence taking}\,\, T_1 \,\, \text{to}\,\, T_2\}.$
\end{mydef}
Since $\delta$ is a finite sum of positive definite distance functions, itself is a positive definite distance function as well.\\
The tree edit operations give rise to a mapping that is a (equivalent) graphical representation of what edit operations apply to each node in the two trees. Let $T_1$ and $T_2$ be two labeled ordered trees with $N_1$ and $N_2$ nodes, respectively. Fix a traversal rule (e.g. postorder to preorder), we define $T[i]$ as the $i^{\text{th}}$ node of $T$ in the traversal. Now with this traversal fixed, we can identify the node $T[i]$ with the number $i$. 

\begin{mydef}[Mapping Between Two Trees]\label{def:match associated with edit script}
A \emph{mapping} between $T_1$ and $T_2$ is a triple $(M, T_1, T_2)$, where $M$ is any set of pair of integers $(i, j)$ where $1 \leq i \leq N_1$, $1 \leq j \leq N_2$, such that for any pair of $(i_1, j_1)$ and $(i_2, j_2)$ in $M$:
\begin{enumerate}
\item[(1)] $i_1 = i_2$ if and only if $j_1 = j_2$. This is called the \emph{one-to-one} condition. 
\item[(2)] $T_1[i_1]$ is to the left of $T_1[i_2]$ if and only if $T_2[j_1]$ is to the left of $T_2[j_2]$. This is called the \emph{sibling order} condition.
\item[(3)] $T_1[i_1]$ is an ancestor of $T_1[i_2]$ if and only if $T_2[j_1]$ is an ancestor of $T_2[j_2]$. This is called the \emph{ancestor order} condition. 
\end{enumerate}
\end{mydef}

$M$ can be viewed as an order (sibling order and ancestor order) preserving mapping taking (a subset of) vertices from one tree to that of another tree. We say that a node is \emph{not touched} if it does not appear as either one of the vertices in the domain of $M$. If we draw a line connecting the two nodes in each pairs that lie in the domain of $M$, then nodes that are not touched do not have lines coming in or going out. Let $I$ and $J$ be the sets of nodes in $T_1$ and $T_2$ respective representing those nodes that are not touched.
\begin{mydef}\label{def:cost_of_mapping}
The cost of such a mapping, with $\gamma$ previously given as above, is defined as:
\begin{equation}
\gamma(M): = \sum_{(i, j)\in M} \gamma(T_1[i], T_2[j]) + \sum_{i\in I}\gamma(i, \lambda) + \sum_{j\in J}\gamma(\lambda, j).
\end{equation}
\end{mydef}
Then it is easy to show that 
\begin{lemma}\label{lemma:edit distance equals minimal cost of mappings}
Given an edit operation sequence $S$ from $T_1$ to $T_2$, there exists mapping $M$ from $T_1$ to $T_2$ such that $\gamma(M) \leq \gamma(S)$. Conversely, for any mapping $M$, there exists a sequence of editing operations such that $\gamma(S) = \gamma(M)$. Therefore:
\begin{equation}
\delta(T_1, T_2) = \min\{\gamma(M)| M\,\, \text{is a mapping from}\,\, T_1\,\, \text{to}\,\, T_2\}.
\end{equation}
\end{lemma}
The above lemma implies that in order to compute the edit distance, it suffices to understand all mappings that are order preserving. Therefore in the following, we will switch our perspective from \emph{transforming} one tree to another, to \emph{mapping} the nodes of one tree to the nodes of anther. To do that, we need some terminologies:
\begin{enumerate}
\item From now on we fix the left-to-right \emph{postorder} traversal rule unless otherwise specified. This rule defines a numbering among all the nodes of a tree $T$. 

\item Let $T[i]$ denote the $i^{\text{th}}$ node (so that we can identity $T[i]$ with $i$), and $l(i)$ be the \emph{number} of the leftmost leaf descendant of the subtree rooted at $T[i]$. Hence when $T[i]$ is a leaf, $l(i) = i$.

\item The parent of $T[i]$ is denoted $p(i)$, and $anc(i)$ denotes the ancestors of $T[i]$, and $desc(i)$ the descendants of $T[i]$. 

\item Let $forest(i..j): = T[i..j]$ be the ordered \emph{subforest} of $T$ induced by the nodes numbered from $i$ to $j$ inclusively. If $i > j$, then $T[i..j] = \emptyset$. 

\item In particular, $T[1..i]$ will be referred to as $forest(i)$, when the tree $T$ is clear in context. Note that $T[l(i)..i]$ is simply the subtree rooted at $T[i]$, and thus will be referred as $tree(i)$.

\item We use $Size(i)$ to denote the number of nodes in $tree(i)$.
\end{enumerate}

\section{Zhang and Shasha's Algorithm Part II: Recurrences}
\label{sec:zhang and shasha's algorithm}
Zhang and Shasha idea of computing $\delta(T_1, T_2)$ is the following:\\

\emph{We use dynamic programming starting from the distances between smaller components of $T_1$ and $T_2$, and build up from that. Portions of a tree is in general a forest, thus it is important to understand the distances between such forests. We build up the tree from the right most node in the postorder traversal in a bottom-up fashion.}\\
\begin{mydef}
We define 
\[forestdist(i'..i, j'..j): = forestdist(T_1[i'..i], T_2[j'..j]): = \delta(T_1[i'..i], T_2[j'..j]),\]
where $\delta$ is defined same as before. And 
\[forestdist(i, j): = forestdist(1..i, 1..j).\] 
Finally, the distance between the subtrees rooted at $i$ and $j$ respectively is denoted as 
\[treedist(i, j) = forestdist(l(i)..i, l(j)..j).\]
\end{mydef}
The dynamic programming algorithm design is based on the following key recursions:
\begin{theorem}[\cite{ZhangShasha1989}]\label{thm:tree_edit_dist_recursion}
For any $i\in desc(i_1)$ and $j\in desc(j_1)$\footnote{Here we have identified $i_1$ with the $i_1^{\text{th}}$ node in $T_1$, given the postorder numbering. Same for $i, j, j_1$.}, then:
\[
forestdist(l(i_1)..i, l(j_1)..j) = \min
\begin{cases}
forestdist(l(i_1)..i-1, l(j_1)..j) + \gamma(T_1[i]\rightarrow \lambda) \\
forestdist(l(i_1)..i, l(j_1)..j-1) + \gamma(\lambda\rightarrow T_2[j]) \\
forestdist(l(i_1)..l(i)-1, l(j_1)..l(j)-1) \\
+ forestdist(l(i)..i-1, l(j)..j-1) + \gamma(T_1[i]\rightarrow T_2[j])
\end{cases}
\]
\end{theorem}
\begin{proof}
First note that since $i\in desc(i_1)$, $l(i_1) \leq i \leq i_1$. Similarly, $l(j_1) \leq j \leq j_1$. To prove this claim, it suffices to find a minimum-cost map between $forest(l(i_1)..i)$ and $forest(l(j_1)..j)$. Notice that $i$ and $j$ are the rightmost nodes of the two forests respectively, and there are three possible configurations of $i$ and $j$ in any mapping $M$:
\begin{enumerate}
\item $i$ is not touched by a line in $M$. Then $(i, \lambda) \in M$. Thus
\[forestdist(l(i_1)..i, l(j_1)..j) = forestdist(l(i_1)..i-1, l(j_1)..j) + \gamma(T_1[i]\rightarrow \lambda).\]

\item $j$ is not touch by a line in $M$. Then similar as above,
\[forestdist(l(i_1)..i, l(j_1)..j) = forestdist(l(i_1)..i, l(j_1)..j-1) + \gamma(\lambda\rightarrow T_2[j]).\]

\item Both $i$ and $j$ are touched by a line (see Figure \ref{fig:linear gap recurrence}). This is the only non-trivial case, and we claim that in this case, $(i, j)\in M$, i.e., $i$ must be mapped to $j$. To prove the claim, we suppose the contrary: suppose $(i, k)$ and $(h, j)$ are in $M$, and $h \neq i, k \neq j$. Thus either $l(l_1) \leq h \leq l(i) - 1$, or $l(i) \leq h \leq i - 1$. The first case implies that $i$ is to the right of $h$, so $k$ must be the right of $j$ by the sibling condition on $M$. But there is no such $k$ in $forest(l(j_1)..j)$. Contradiction, and this forces $l(i) \leq h \leq i - 1$, i.e., $i$ is a proper ancestor of $h$. By the ancestor condition, $k$ is a proper ancestor of $j$, which is again impossible in $forest(l(j_1)..j)$. Therefore $h = i$. By a symmetric argument, $k = j$ as well. Therefore $i$ is mapped to $j$, and by the ancestor condition on $M$, the subtree rooted at $i$ must be mapped into the subtree rooted at $j$. Therefore the last case follows:
\begin{align*}
forestdist(l(i_1)..i, l(j_1)..j) & = forestdist(l(i_1)..l(i)-1, l(j_1)..l(j)-1) \\
& + forestdist(l(i)..i-1, l(j)..j-1) + \gamma(T_1[i]\rightarrow T_2[j]).
\end{align*}

\begin{figure}[!htb]
\centering
\includegraphics[scale = .45]{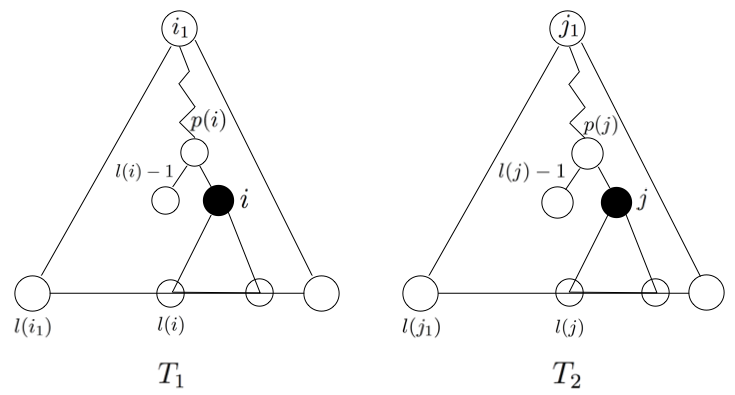}
\caption{Both $i$ and $j$ are touched by a line. In this case, $i$ must be mapped to $j$.}
\label{fig:linear gap recurrence}
\end{figure}

\end{enumerate}
\end{proof}

Theorem \ref{thm:tree_edit_dist_recursion} has the following corollary:
\begin{coro}[\cite{ZhangShasha1989}]\label{coro:tree_edit_dist_recursion}
For any $i\in desc(i_1), j\in desc(j_1)$, 
\begin{enumerate}
\item If $l(i) = l(i_1)$, and $l(j) = l(j_1)$, i.e., $i$ is on the path from $i_1$ to its leftmost leaf $l(i_1)$, and $j$ is on the path from $j_1$ to leftmost leaf $l(j_1)$, then
\begin{align*}
treedist(i, j) & = forestdist(l(i_1)..i, l(j_1)..j) \\
& = \min
\begin{cases}
forestdist(l(i_1)..i-1, l(j_1)..j) + \gamma(T_1[i]\rightarrow \lambda) \\
forestdist(l(i_1)..i, l(j_1)..j-1) + \gamma(\lambda\rightarrow T_2[j]) \\
forestdist(l(i)..i-1, l(j)..j-1) + \gamma(T_1[i]\rightarrow T_2[j])
\end{cases}
\end{align*}

\item If $l(i)\neq l(i_1)$, or $l(j)\neq l(j_1)$, 
\[
forestdist(l(i_1)..i, l(j_1)..j) = \min
\begin{cases}
forestdist(l(i_1)..i-1, l(j_1)..j) + \gamma(T_1[i]\rightarrow \lambda) \\
forestdist(l(i_1)..i, l(j_1)..j-1) + \gamma(\lambda\rightarrow T_2[j]) \\
forestdist(l(i_1)..l(i)-1, l(j_1)..l(j)-1) + treedist(i, j)
\end{cases}
\]
\end{enumerate}
\end{coro}

\begin{proof}
The first part is easy: if $l(i) = l(i_1)$, and $l(j) = l(j_1)$, then $T[l(i_1), i] = T[l(i)..i] = tree(i)$. Similarly $T[l(j_1)..j] = T[l(j)..j] = tree(j)$, thus the first equality in part (1) follows. The rest of part (1) follows from the fact that $forestdist(l(i_1)..l(i) - 1, l(j_1)..l(j) - 1) = forestdist(\emptyset, \emptyset) = 0$.\\
For part (2), note that
\[forestdist(l(i_1)..i, l(j_1)..j) \leq forestdist(l(i_1)..l(i) - 1, l(j_1)..l(j) - 1) + treedist(i, j),\]
since the latter formula represents a particular (and therefore possibly suboptimal) mapping of $forest(l(i_1)..i)$ to $forest(l(j_1)..j)$. For the same reason,
\[treedist(i, j) \leq forestdist(l(i)..i-1, l(j)..j) + \gamma(i\rightarrow j).\]
Therefore $forestdist(l(i_1)..l(i) - 1, l(j_1)..l(j) - 1) + treedist(i, j)$ is a tighter upper bound for $forestdist(l(i_1)..i, l(j_1)..j)$. Since we are looking for the minimum value of $forestdist(l(i_1)..i, l(j_1)..j)$, we can use a tighter upper bound with affecting the result. 
\end{proof}

The above theorem and corollary serve as the basis for using dynamic programming to compute tree edit distance. More precisely, Theorem \ref{thm:tree_edit_dist_recursion} implies that in order to compute $treedist(i_1, j_1) = forestdist(l(l_1)..i_1, l(j_1)..j)$\footnote{pick $i = i_1$ and $j = j_1$ in the above.}, we need in advance almost all values of $treedist(i, j)$ for $i\in desc(i_1)$, and $j\in desc(j_1)$, as long as $l(i) \neq l(i_{1})$ or $l(j) \neq l(j_{1})$. This suggests a bottom up approach for computing $treedist(i_1, j_1)$: \\

Compute $treedist(i, j)$, for $i = l(i_1), \cdots, i_1$, and $j = l(j_1), \cdots, j_1$. The number of all such pairs is on the order of $N_{1}^{2} N_{2}^{2}$, where $N_1$ is the number of nodes in $tree(i_1)$, and $N_2$ is the number of nodes in $tree(j_1)$.\\

However, Corollary \ref{coro:tree_edit_dist_recursion} suggests that we don't have to compute all such intermediate distance of subtree pairs. Given two subtrees $tree(i)$ and $tree(j)$, to actually compute the distance $treedist(i, j)$, we need the distance between all the \emph{prefixes} of the two subtrees, where a prefix of a tree is the result of deleting the rightmost node in the postorder numbering. We can keep deleting the rightmost nodes to get all the prefixes. \\Now if $i$ is in the path from $l(i_1)$ to $i_1$, and $j$ is in the path from $l(j_1)$ to $j$, then in computing $treedist(i_1, j_1)$, we get $treedist(i, j)$ as a byproduct, since $tree(i)$ and $tree(j)$ are prefixes of $tree(i_1)$ and $tree(j_1)$, respectively. Thus the upshot is this:
\emph{In computing the distance of subtree pairs, we can skip those pair in which each subtree is the prefix of some super tree whose root is an ancestor of the root of this subtree}. It is easy to see that those are exactly subtrees rooted at nodes that do \emph{not} have a left sibling. This motivates the following definition:
\begin{mydef}[\cite{ZhangShasha1989}]\label{def:key_roots}
Given a tree $T$, we define the set of \emph{LR key roots} of $T$ to be the union of the root of $T$, together with all nodes that have left siblings. Here \emph{LR} refers to the left-to-right postorder numbering. 
\end{mydef}
Therefore all we need to compute are distances between pairs of subtrees rooted at LR key roots. We formulate the pseudocode as follows (see Algorithm \ref{alg:tree_edit_distance_dynamic}):

\begin{algorithm}
\caption{Dynamic Programming Algorithm for Tree Edit Distance}\label{alg:tree_edit_distance_dynamic}
\label{Fingerprint}
\begin{algorithmic}
\State Input: Tree $T_1$ and $T_2$
\State Output: $treedist(i, j)$, where $1 \leq i \leq |T_1|$, and $1 \leq j \leq |T_2|$
\State Preprocessing: Compute the $l$ function, and the LR key roots for $T_1$ and $T_2$, put them in the array $KR_1$ and $KR_2$ respectively
\For {($i' = 1 \to KR_1.size()$)}
    \For{($j' = 1 \to KR_2.size()$)}
        \State $i = KR_1[i']$; \Comment{(pick the i'th key root)}
        \State $j = KR_2[j']$; \Comment{(pick the j'th key root)}
        \State Compute $treedist(i, j)$ using $forestdist(i', j')$ for $1 \leq i' \leq i, 1 \leq j' \leq j$;
    \EndFor
\EndFor
\end{algorithmic}
\end{algorithm}

To compute each $treedist(i, j)$, the $forestdist$ values computed and used here are put in a temporary array that is freed once the corresponding $treedist$ is computed. The $treedist$ values are put in the permanent $treedist$ array. The computation of $tree(i, j)$ is again bottom-up: starting from the smallest prefixes of $tree(i)$ and $tree(j)$ and build up. The details are given in \cite{ZhangShasha1989}, p.1253.

\section{Zhang and Shasha's Algorithm Part III: Algorithm Complexity Analysis}
\label{sec:zhang and shasha algorithm complexity analysis}
We first bound the size of the set of key roots in a tree. 
\begin{lemma}[\cite{ZhangShasha1989}]\label{lemma:key_root_size_bound}
The set of LR key roots of a tree $T$ is less than or equal to the number of leaves of $T$.
\end{lemma}
\begin{proof}
We show that for distinct key roots $i$ and $j$ have distinct leftmost leaf descendants $l(i)$ and $l(j)$, respectively, thereby proving the claim. Suppose not, and without loss of generality assume that $i < j$. Then $i$ is on the path from $l(j)$ to $j$. From the definition of $l(j)$, $i$ does not have any left siblings, contradicting the assumption that $i$ is a key root. Therefore $l(i) \neq l(j)$.
\end{proof}

The complexity of the above algorithm is rooted in the number of pairs of subtrees whose distance are being computed. For any node $i$ in $T$, we say that it participates the algorithm computation if it belongs to such a subtree, rooted at a key root. Then it is easy to see that the number of times any given node participates equals the number of its key root ancestors. We define the quantity to be the \emph{collapsed depty} of $i$:
\begin{mydef}[\cite{ZhangShasha1989}]\label{def:cdepth}
The collapsed depth of a node $i$, denoted as $cdepth(i)$, is given by the number of the key root ancestors of $i$, including $i$ if it is a key root. And we set $cdepty(T): = \displaystyle\max_{i \in V(T)} cdepth(i)$. 
\end{mydef}
Then by Lemma \ref{lemma:key_root_size_bound}, 
\begin{equation}
cdepth(T) \leq \min (depth(T), |leaves(T)|).
\end{equation}
Now we bound the total number of participating nodes:
\begin{lemma}[\cite{ZhangShasha1989}]\label{lemma:total_number_of_participating_nodes}
Let $K$ be the number of LR key roots of $T$, and $N$ be the number nodes of $T$, then
\begin{equation}\label{equation:total_number_of_participating_nodes}
\sum_{i=1}^{K} Size(i) = \sum_{j = 1}^{N}cdepth(j).
\end{equation}
\end{lemma}
\begin{proof}
Note that the left hand size in $(\ref{equation:total_number_of_participating_nodes})$ is exact the total number of participating nodes, counted with multiplicity, in the computation of the tree edit distance as one of the trees. Note that each participating node $i$ is counted $cdepth(i)$ times in the left summation. Moreover, each node $j$ such that $cdepth(j) > 0$ is participating. Therefore the two summations agree. 
\end{proof}

Now we are in a position to bound the running time and space usage of the algorithm \ref{alg:tree_edit_distance_dynamic}:
\begin{theorem}[\cite{ZhangShasha1989}]\label{thn:tree_edit_distance_analysis}
The above algorithm in computing the edit distance between $T_1$ and $T_2$ takes time 
\begin{equation}
O\Big(|T_1||T_2|\cdot \min(depth(T_1), |leaves(T_1)|)  \cdot  \min(depth(T_2), |leaves(T_2)|)\Big),
\end{equation}
and space 
\begin{equation}
O\Big(|T_1| |T_2|\Big).
\end{equation}
\end{theorem}

\begin{proof}
For the space complexity, we use an array to keep the key roots, $treedist$ values and $forestdist$ values, each takes $O(|T_1||T_2|)$ space.\\
For the time complexity, the preprocessing takes linear time in computing $l$ and the key roots. In the main loop, we are computing $treedist(i, j)$ for each $1 \leq i \leq K_1$, and $1 \leq j \leq K_2$, where $K_1$ and $K_2$ are the size of the LR key roots of $T_1$ and $T_2$ respectively. $treedist(i, j)$ takes time $Size(i)\cdot Size(j)$, since that's the number of pairs of all prefixes of $tree(i)$ and $tree(j)$. Therefore the running time after the preprocessing is:
\begin{align*}
\sum_{i = 1}^{K_1} \sum_{j = 1}^{K_2} Size(i)\cdot Size(j) & = \left(\sum_{i = 1}^{K_1} Size(i)\right) \left(\sum_{j = 1}^{K_2} Size(j)\right)\\
& = \left(\sum_{i = 1}^{N_1} cdept(i)\right) \left(\sum_{j = 1}^{N_2} cdept(j)\right) \tag{By Lemma \ref{lemma:total_number_of_participating_nodes}}\\
& \leq |T_1||T_2|cdepth(T_1)\cdot cdepth(T_2)\\
& \leq |T_1||T_2|\cdot \min(depth(T_1), |leaves(T_1)|)  \cdot  \min(depth(T_2), |leaves(T_2)|
\end{align*}
where the last inequality follows from Lemma \ref{lemma:key_root_size_bound}. This concludes the proof of the theorem. 
\end{proof} % A regular chapter, starts with '\chapter{Title}'
\chapter{Tree Edit Distance with Gaps}
\label{chap:tree edit distance with gaps}

In this chapter, we study edit distance between trees with gaps, in particular, gap models, and gap cost functions. The classic edit distance can be viewed as gaped edit distance with linear gap costs. 

In Section \ref{sec:motivation for gaps}, we discuss motivations for introducing gaps in comparing tree similarities. In Section \ref{sec:general gap model with arbitrary convex gap penalty function}, we compute the general gap edit distance with affine gap costs between two binary trees of size $m$ and $n$ respectively in $O(m^{3}n^{2} + m^{2} n^{3})$ time. The computation of this distance between general trees is shown to be NP-hard (see \cite{Touzet2003}). 

In Section \ref{sec:complete subtree model with affine gap cost function}, we study the complete subtree gap model, which is a weaker model first proposed by Touzet \cite{Touzet2003}. We present an algorithm that computes the corresponding edit distance with affine gap costs in $O(m^{2}n^{2})$ time. In Section \ref{sec:application to contour tree comparisons}, we discuss an application of the complete subtree gap model to contour tree comparisons. Finally in Section \ref{sec:further improvements}, some further improvements are discussed.

We assume that all trees considered in this chapter are ordered and labeled with characters taken from a finite alphabet $\Sigma$. We use $\perp$ to denote a special characters outside $\Sigma$. 

\section{Motivations and Main Results}
\label{sec:motivation for gaps}

Recall the one main motivation for studying tree similarity comparison is that many (complicated) geometric shapes have (simpler) underlying tree structures that capture some key topological or geometric properties of the original shapes. Thus, the problem of shape comparison can be reduced to tree comparison. However, in many applications, such geometric shapes often have noise present in their input, which often get reflected in the underlying tree structures. Therefore, it is desirable to delete such ``auxiliary'' portions in trees, which do not represent the true topology or geometry of the original shapes, before comparing them. 

The classic tree edit distance allows pointwise deletion, i.e. one node at a time. There are two natural generalizations. First, in addition to pointwise insertion or deletion, multiple nodes could be inserted or deleted. Second, instead of charging every deleted node equally, more general cost functions could be used. 

In this more general version of tree edit distance, nodes can be inserted or deleted in groups, called \emph{gaps}, which is analogous to gaps in sequence alignment (see Definition \ref{def:gaps in sequence alignment}). Moreover, we can consider affine (or more generally convex) functions for gap costs. This is again motivated by the fact that in some applications, it is more probable to have a ``big'' noise than several ``small'' noise scattered in the input.

What is a good model (e.g. intrinsic and computable) for gaps in trees? One natural definition of a gap is a \emph{connected component} of the nodes deleted. This is analogous to the sequence alignment case, in which gaps are \emph{largest consecutive nodes deleted}. This gap model, referred to as the \emph{general gap model} in this thesis, was first proposed by Touzet \cite{Touzet2003}, whose motivation at the time came from the problem of comparing secondary structures of RNA. 

Unfortunately, Touzet \cite{Touzet2003} showed that, even with affine gap cost functions, the computation of this general gap edit distance is NP-hard\footnote{More precisely, the decision problem: \emph{given two ordered labeled trees and a positive integer $k$, decide wether the general gap edit distance is bounded from above by $k$}, is NP-hard.}. For this reason, gapped edit distance with nonlinear gap cost functions has received fewer studies than the classic edit distance, which can be viewed as gapped edit distance with linear gap costs (despited of the choice of gap models). Rolf Backofen et al \cite{Backofen2007} studied the application of edit distance with gaps in RNA comparison.  S. Schirmer and R. Giegerich \cite{Schiremeretal2011} studied tree alignment with affine gaps that concerns the problem of optimal embedding two trees into a common tree, first proposed by T. Jiang, L. Wang and K. Zhang in \cite{JiangWangzhang1995}. G. Blin and H. Touzet \cite{BlinTouzet2006} studied the application of tree alignment in computational biology. 

Here is the central question we consider in this thesis: Even though computing the general gap tree edit distance is NP-hard, is it possible to weaken this distance and get a computable measure of similarity between two trees? In the following, two ways to weaken the general gap edit distance are considered. We could either compare more specific trees (e.g. binary trees), or use a more restrictive gap model (e.g. complete subtree model). It turns out that both of these two approaches yield polynomially computable distances.

%%%%%%%
\section{General Gap Tree Edit Distance Between Binary Trees}
\label{sec:general gap model with arbitrary convex gap penalty function}

\subsection{Genera Gap Model, Edit Distance and Mapping}
\begin{figure}[!htb]
\centering
\includegraphics[scale = .5]{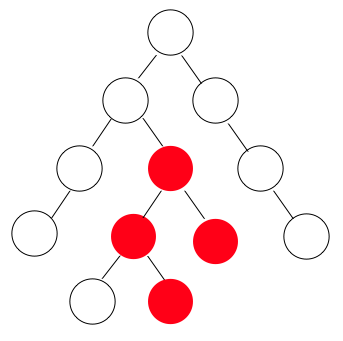}
\caption{Red nodes in this binary tree forms a gap.}
\label{fig:gaps in binary tree}
\end{figure}

\begin{mydef}[General Gaps Model, \cite{Touzet2003}]\label{def:general gap model}
Given an ordered labeled tree $T$ with vertex set $V$ and edge set $E$. A gap $g$ is a tree with vertex set a subset of $V$ and edges in $E$ whose both end points lie in that subset. A node in $g$ is called a gap node. Topologically, a gap is a subtree of $T$ (see Figure \ref{fig:gaps in binary tree}).
\end{mydef}

The corresponding edit operations as:
\begin{mydef}[Edit Operations]\label{def:formal admissible edit operations}
Here are the tree type of edit operations in the general gap model:

$1.$ Relabel a node;

$2.$ Delete a gap, and descendants of a gap will become children of the parent of the root of the gap;

$3.$ Insert a gap.
\end{mydef}

Note that inserting (resp. deleting) a gap in one tree corresponds to deleting (resp. inserting) a gap in the other. 

Each edit operation has a nonnegative cost. Given two trees $T_{1}, T_{2}$ with vertex set $V_{1}$ and $V_{2}$, respectively. For the cost of relabeling, choose a metric (symmetric, positive definite and satisfies the triangle inequality)
\begin{equation}\label{def:cost function for relabeling}
p: V_{1} \times V_{2} \longrightarrow \RR^{\geq 0}.
\end{equation}
Then $p(u, v)$ defines the cost of changing the label on $u\in V_{1}$ to that on $v\in V_{2}$.
For the cost of deletion and insertion of gaps, first consider an arbitrary function
\begin{equation}\label{def:arbitrary cost function for gaps}
w: \ZZ^{+}\longrightarrow \RR^{\geq 0}.
\end{equation}
such that the cost of deleting or inserting a gap $g$ is given by $w(|g|)$, where $|g|$ denotes the number of nodes in $g$. Thus, the cost of $g$ only depends on the size of $g$. One could generalize this even further by considering functions depending on other properties of $g$ (e.g. height, total degree, etc).

Based on the heuristic that a large gap is more likely to occur than multiple isolated small gaps, convex gap cost functions are more suitable for our considerations:
\begin{equation}\label{def:convex gap cost function}
w: \ZZ^{+}\longrightarrow \RR^{\geq 0}, \quad w(k_{1} + k_{2}) \leq w(k_{1}) + w(k_{2}), \quad \forall\, k_{1}, k_{2}\in \ZZ^{+}.
\end{equation}
for gap penalty. In particular, an affine function
\begin{equation}\label{def:affine gap cost function in edit distance}
w(k) = 
\begin{cases}
0 & \text{for}\,\, k = 0\\
a + bk & \text{for}\,\, k \in \ZZ^{+}, a\geq 0, b > 0
\end{cases}
\end{equation}
is a convex function (see Lemma \ref{lemma:affine functions are convex}). In general, the more complicated the gap cost function is, the more difficult the computations will be. In the following, we assume that all gap costs are affine unless otherwise specified. 

\begin{mydef}[Edit Script]\label{def:edit script and its cost in general gap model and arbitrary gap cost function}
Given two trees $T_{1}$ and $T_{2}$. An edit script $S$ from $T_{1}$ to $T_{2}$ is a sequence of edit operations $S = \{S_{1}, S_{2}, \cdots, S_{n}\}$ that transforms $T_{1}$ to $T_{2}$. The cost of $S$ is defined to be
\begin{equation}
C(S): = \sum_{i = 1}^{n}C(S_{i}),
\end{equation}
where $C(S_{i})$ is the cost of the $i$th edit operation. 
\end{mydef}

\begin{mydef}[General Gap Tree Edit Distance]
\label{def:edit distance with general gap model and arbitrary gap cost function}
Given two ordered labeled trees $T_{1}$ and $T_{2}$. The general gap edit distance between $T_{1}$ and $T_{2}$ is defined to be
\begin{equation}\label{equ:edit distance with general gap model and arbitrary gap cost function}
\gamma(T_{1}, T_{2}): = \min\{C(S)|S\,\,\text{is an edit script taking $T_{1}$ to $T_{2}$}\}.
\end{equation}
\end{mydef}

A mapping between two trees, which is a graphical representation of an edit script, can be defined in exactly the same fashion as in classic edit distances (\ref{def:match associated with edit script}):
\begin{mydef}[Mapping Between Two Trees]\label{def:mapping between two trees}
Given two trees $T_{1}$ and $T_{2}$. A mapping between $T_1$ and $T_2$ is a triple $(M, T_1, T_2)$, where $M$ is a subset of $V_{1}\times V_{2}$, such that for any $(u, v)$ and $(u', v')$ in $M$:
\begin{enumerate}
\item[(1)] $u = u'$ if and only if $v = v'$, called the \emph{one-to-one} condition. 
\item[(2)] $u$ is to the left of $u'$ if and only if $v$ is to the left of $v'$, called the \emph{sibling order} condition.
\item[(3)] $u$ is an ancestor of $u'$ if and only if $v$ is an ancestor of $v'$, called the \emph{ancestor order} condition. 
\end{enumerate}

Given a mapping $M$, define its cost to be:
\begin{equation}\label{def:cost of mapping in general gap model with affine gap cost function}
C(M): = \sum_{(u, v) \in M} p(u, v) + \sum_{g\in G}a + b |g|, \quad u\in V_{1}\,\, v\in V_{2},
\end{equation}
where $G$ is the set of all gaps in $M$.

\end{mydef}

By Lemma \ref{lemma:edit distance equals minimal cost of mappings}, we still have 
\[\gamma(T_{1}, T_{2}) = \gamma(M): = \min\{C(M) | M\,\,\text{is a mapping from $T_{1}$ to $T_{2}$}\}.\]
Therefore computing the edit distance is equivalent to computing the minimal cost mapping.

\subsection{Binary Tree Case}
Since the computation of the general gap edit distance is NP-hard for arbitrary trees \cite{Touzet2003}, we compute this distance for \emph{binary} trees. We prove the following main theorem of this thesis:

\begin{theorem}[Main Theorem]
\label{thm:general gap edit distance between binary trees with affine gap penalty}

Given two ordered labeled binary trees $T_{1}$ and $T_{2}$ with vertex set $V_{1}: = V(T_{1})$ and $V_{2}: = V(T_{2})$ respectively, and an affine gap cost function. Let $m := |V_{1}|$ and $n := |V_{2}|$. The general gap edit distance between $T_{1}$ and $T_{2}$ can be computed in $O(m^{3}n^{2} + m^{2}n^{3})$ time. If $m \geq n$, then the running time is $O(m^{5})$. 
\end{theorem}

We use a dynamic programming approach to prove this theorem, similar to Zhang and Shasha's approach \cite{ZhangShasha1989} in the classic edit distance case. Given a matching $M$ and a pair of nodes $(u, v)\in V_{1} \times V_{2}$. There are three possibilities: either $u$ is matched to $v$; or $u$ is a gap node; or $v$ is gap node. Since the gap cost function is affine, the penalty for starting a gap is different from that for continuing a gap. Moreover, a gap node $u$ is continuing a gap if and only if its parent node, denoted as $p(u)$, is a gap node. Thus, to determine whether a gap node is starting or continuing a gap, we need the information about its parent node. This suggests that we order the nodes according to \emph{preorder} traversals as apposed to postorder traversals in the classic edit distance.

Order all the nodes in $T_{1}$ and $T_{2}$ via preorder traversal and enumerate the nodes in $T_{1}$ as $1, 2, \cdots, m$, and the nodes in $T_{2}$ as $1, 2, \cdots, n$. Identify $T_{1}[i]$, the $i$th node, with its index $i$, $i = 1, \cdots m$. Same for $T_{2}[j]$, $j = 1, \cdots, n$. Let $T_{1}[i'..i]$ and $T_{2}[j'..j]$ be the subforests defined in Section \ref{sec:setups and terminologies}, and $\forestdist(T_{1}[i'..i], T_{2}[j'..j])$ be the edit distance between $T_{1}[i'..i]$ and $T_{2}[j'..j]$. Define three auxiliary functions:
\begin{mydef}\label{def:auxiliary functions in binary tree comparison in general gap model}
For $1\leq i' \leq i \leq m$, and $1 \leq j' \leq j \leq n$, set:
\begin{equation}
\begin{cases}
Q[i'..i, j'..j] := \forestdist(T_{1}[i'..i], T_{2}[j'..j]);\\
Q_{\perp *}[i'..i, j'..j] :=  \forestdist(T_{1}[i'..i], T_{2}[j'..j])\,\, \text{such that $i$ is a gap point};\\
Q_{*\perp}[i'..i, j'..j] := \forestdist(T_{1}[i'..i], T_{2}[j'..j])\,\, \text{such that $j$ is a gap point}
\end{cases}
\end{equation}
\end{mydef}
With this definition 
\[\gamma(T_{1}, T_{2}) = Q[1..m, 1..n].\]

We define the boundary conditions of the auxiliary functions as follows. Note first that $T[i..i'] = T[\emptyset]$ if $i' < i$. Set 

\begin{align}
Q[\emptyset, \emptyset] = 0,\notag\\
Q[1..i, \emptyset] = \infty, \tag{for $1 \leq i \leq m$}\notag\\
Q[\emptyset, 1..j] = \infty, \tag{for $1 \leq j \leq n$}\label{equ:boundary condition for Q}
\end{align}

Moreover set
\begin{align}
Q_{\perp *}[1..i, \emptyset] = a + bi, \tag{for $1 \leq i \leq m$}\notag\\
Q_{\perp *}[\emptyset..1..j] = \infty, \tag{for $1 \leq j \leq n$} \label{equ:boundary condition for Qperpstar}
\end{align}
since it is impossible to match an empty tree with $T_{2}[1..j]$ such that the formal ends with a gap node; and there is a unique matching between $T_{1}[1..i]$ with an empty tree: we have $i$ gap points.

By symmetry, set
\begin{align}
Q_{*\perp}[1..i, \emptyset] = \infty, \tag{for $1 \leq i \leq m$} \notag\\
Q_{*\perp}[\emptyset, 1..j] = a + bj, \tag{for $1 \leq j \leq n$}\label{equ:boundary condition for Qstartperp}
\end{align}

\begin{theorem}[Recurrence of Auxiliary Matrices in General Gap Model for Binary Trees]
\label{thm:recurrence for aux functions in binary tree case}
Given the preorder ordering on the nodes of two ordered labeled trees $T_{1}$ and $T_{2}$. Fix nodes $i_{1}\in V_{1}, j_{1}\in V_{2}$. For any $i\in \desc(i_{1})$ and $j\in \desc(j_{1})$, we have the following recurrence relations:
\begin{equation}\label{equ:recurrence for Q}
Q[i_{1}..i, j_{1}..j] = \min 
\begin{cases}
Q[i_{1}..i-1, j_{1}..j-1] + p(i, j)\\
Q_{\perp *}[i_{1}..i, j_{1}..j]\\
Q_{*\perp}[i_{1}..i, j_{1}..j] 
\end{cases}
\end{equation}
%\vspace{-7ex}
\begin{equation}\label{equ:recurrence for Qperpstar}
Q_{\perp *}[i_{1}..i, j_{1}..j] = \min
\begin{cases}
Q[i_{1}..i-1, j_{1}..j] + (a + b) \\
Q_{\perp *}[i_{1}..i-1, j_{1}..j] + b \\
\displaystyle\min_{j_{1}\leq k \leq  j}\big\{Q_{\perp *}[i_{1}..p(i), j_{1}..k] \\
\quad \quad \,\, + Q[p(i)+1..i-1, k+1..j] + b\big\}
\end{cases}
\end{equation}
%\vspace{-5ex}
\begin{equation}\label{equ:recurrence for Qstarperp}
Q_{*\perp}[i_{1}..i, j_{1}..j] = \min
\begin{cases}
Q[i_{1}..i, j_{1}..j-1] + (a + b)\\
Q_{*\perp}[i_{1}..i, j_{1}..j-1] + b \\
\displaystyle\min_{i_{1}\leq k \leq i}\big\{Q_{*\perp}[i_{1}..k, j_{1}..p(j)]\\
\quad \quad \, \,  + Q(k+1..i, p(j)+1..j-1) + b\big\}
\end{cases}
\end{equation}
Here $p(i)$ (resp. $p(j)$) is the index of parent node (if exists) of $i$  (resp. $j$).
\end{theorem}

\begin{proof}
We prove recurrence (\ref{equ:recurrence for Q}) and (\ref{equ:recurrence for Qperpstar}). Recurrence (\ref{equ:recurrence for Qstarperp}) can be obtained by a symmetric argument. We first assume that both $i_{1}$ and $j_{1}$ have nontrivial siblings.

$\bullet$ In the first recurrence of $Q[i_{1}..i, j_{1}..j]$, there are three cases:

(1) None of $i$ or $j$ is a gap point, then $i$ must be matched to $j$, and 
\[Q[i_{1}..i, j_{1}..j] = Q[i_{1}..i-1, j_{1}..j-1]+ p(i,j),\] 
where $p(i, j)$ is the cost of matching $i$ with $j$.

(2) $i$ is a gap point, then 
\[Q[i_{1}..i, j_{1}..j] = Q_{\perp *}[i_{1}..i, j_{1}..j-1].\]

(3) Similarly if $j$ is a gap point, then 
\[Q[i_{1}..i, j_{1}..j] = Q_{*\perp}[i_{1}..i, j_{1}..j].\]

The above exhaust all the possibilities, hence proves (\ref{equ:recurrence for Q}).\\
$\bullet$ Next we prove the second recurrence in which $i$ is a gap point. Let $p(i)$ be the parent of $i$. If $i$ is the root then $p(i): = \emptyset$. For the moment we assume that $i$ has a non-trivial sibling. There are three cases:

(1) If $p(i) = \emptyset$ or $p(i)$ exists and is not a gap node, then $i$ is starting a new gap and hence gets penalized with $a + b$: 
\[Q_{\perp *}[i_{1}..i, j_{1}..j] = Q[i_{1}..i-1, j_{1}..j] + (a + b).\]

(2) If $p(i)$ is a gap node and $i$ is its \emph{left} child (see Figure \ref{fig:general_gap_binary_recurrence_left}). Then $i$ is continuing a preexisting gap, hence only gets penalized with $b$. In the preorder ordering, $p(i) = i-1$, therefore 
\[Q_{\perp *}[i_{1}..i, j_{1}..j] = Q_{\perp *}[i_{1}..i-1, j_{1}..j] + b,\]

\begin{figure}[!htb]
\centering
\includegraphics[scale = .45]{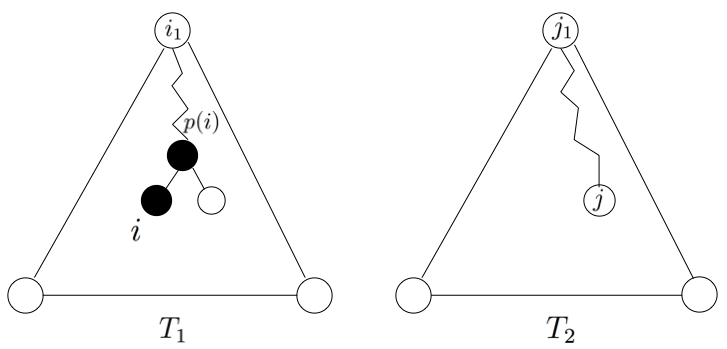}
\caption{$p(i)$ is a gap node and $i$ is its left child. Gap nodes are labeled black.}
\label{fig:general_gap_binary_recurrence_left}
\end{figure}

(3) If $p(i)$ is a gap node and $i$ is its \emph{right} child (see Figure \ref{fig:general_gap_binary_recurrence_right}). Then $i$ is continuing a preexisting gap, hence gets penalized with $b$ as well. Then the left child of $p(i)$ has index $p(i) + 1$. The subforest $T_{1}[i_{1}..p(1)]$ is matched to a subforest $T_{2}[j_{1}..k]$ with $p(1)$ being a gap node, for some $k\in [j_{1}, j]\cap \ZZ$; hence the subtree rooted at $p(i) + 1$ will be matched with the remaining part of $T_{2}[j_{1}..j]$, which is $T_{2}[k+1..j]$. Therefore for $j_{1}\leq k \leq j$,
\[Q_{\perp *}[i_{1}..i, j_{1}..j] = Q_{\perp *}[i_{1}..p(i), j_{1}..k] + Q[p(i) + 1..i-1, k+1..j] + b.\] 

\begin{figure}[!htb]
\centering
\includegraphics[scale = .45]{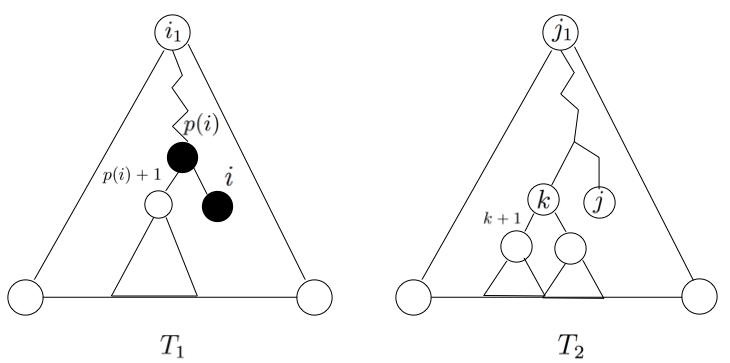}
\caption{$p(i)$ is a gap node and $i$ is its right child. Gap nodes are labeled black.}
\label{fig:general_gap_binary_recurrence_right}
\end{figure}

$\bullet$ To complete the proof, it is left to prove (\ref{equ:recurrence for Qperpstar}) where $i$ is the only child of $p(i)$ since it is immediate that (\ref{equ:recurrence for Q}) still holds in this case. In this case, $p(i) = i-1$ in the preorder ordering. There are two cases:

(1) $p(i)$ is not a gap point, and hence $i$ is starting a new gap:
\[Q_{\perp *}[i_{1}..i, j_{1}..j] = Q[i_{1}..i-1, j_{1}, j] + (a + b).\]

(2) $p(i)$ is a gap point, then $i$ is continuing a preexisting gap:
\[Q_{\perp *}[i_{1}..i, j_{1}..j] = Q_{\perp *}[i_{1}..i-1, j_{1}..j] + b.\]

Notice that $T_{1}[p(i)+1..i-1] = T_{1}[i..i-1] = \emptyset$. Consequently, $Q[p(i)+1..i-1, k+1..j] = \infty$ by the boundary conditions for $Q$ (\ref{equ:boundary condition for Qperpstar}). Therefore (\ref{equ:recurrence for Qperpstar}) still holds in this case.

Combining with the above, we have proved recurrence (\ref{equ:recurrence for Qstarperp}).
\end{proof}

\emph{Algorithm for computing $Q[1..m, 1..n]$}:
Let $\treedist(i, j): = Q[i..r(i), j..r(j)]$, where $r(i)$ is the index of the rightmost leaf descendant of the subtree rooted at $i$. In particular, if $i$ is a leaf, then $r(i) = i$. Thus, the edit distance between $T_{1}$ and $T_{2}$ in the general gap model is $\treedist(m, n)$, and can be computed by algorithm \ref{alg:tree edit distance general gap model dynamic algorithm} below.

\begin{algorithm}
\caption{General Gap Tree Edit Distance}\label{alg:tree edit distance general gap model dynamic algorithm}
\begin{algorithmic}
\State Input: Tree $T_1$ and $T_2$
\State Output: $\treedist(i_{1}, j_{1})$, where $1 \leq i_{1} \leq m: = |T_1|$, and $1 \leq j_{1} \leq n: = |T_2|$
\State Preprocessing: Compute the index of the parent of each node and the $r$ function
\For {($i_{1}, j_{1} = 1$; $i_{1}\leq m, j_{1}\leq n$; $i_{1}$++, $j_{1}$++)}
    	\For{($i = i_{1}$; $i \leq r(i_{1})$;  $i$++)}
		\For{($j = j_{1}$; $j \leq r(j_{1})$;  $j$++)}
        \State Compute $\treedist(i_{1}, j_{1})$ by first compute $Q_{\perp *}[i_{1}..i, j_{1}..j]$,\\ \quad \quad \quad \quad \, then compute $Q_{* \perp}[i_{1}..i, j_{1}..j]$
        		\EndFor
	\EndFor
\EndFor
\end{algorithmic}
\end{algorithm}

\emph{Running Time Analysis}:
A crude upper bound for the time to compute the above recurrences can be computed as follows. We can preprocess $r(i)$ and $r(j)$ for each $i\in V_{1}, j\in V_{2}$. Each computation can be done in linear time. The total time needed to compute $Q[l(i_{1})..i, l(j_{1})..j]$, $Q_{\perp *}[l(l_{1})..i, l(j_{1})..j]$ and $Q_{*\perp}[l(_{1})..i, l(j_{1})..j]$ is upper bounded by
\[3 + 2 + m + 2 + n = 7 + m + n.\]
Since there are $O(m^{2}n^{2})$ many subforest $T[l(i_{1})..i, l(j_{1})..j]$, the upper bound  for computing all of these recurrence is:
\[(7 + (m+n))m^{2}n^{2} = O(m^{3}n^{2} + m^{2}n^{3}).\]
Hence Theorem \ref{thm:general gap edit distance between binary trees with affine gap penalty} is proved.

%%%%%%%
\section{Complete Subtree Gap Tree Edit Distance}
\label{sec:complete subtree model with affine gap cost function}

In this section, we study a weaker model of gaps, first proposed by Touzet:
\begin{mydef}[Complete Subtree Gap Model, \cite{Touzet2003}]
\label{def:complete subtree gap model}
Given a tree $T$ with vertex set $V$. A gap $g_{v}$ of $T$ is the complete subtree rooted at some vertex $v\in V$.
\end{mydef}

\begin{figure}
\centering
\begin{minipage}{.5\textwidth}
  \centering
  \includegraphics[width=.95\linewidth]{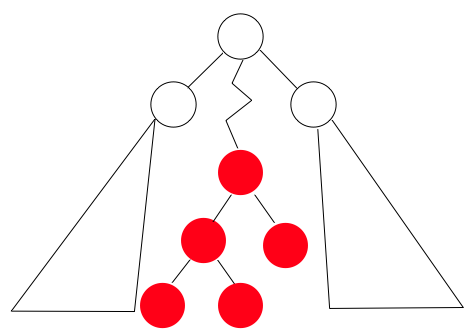}
  %\captionof{figure}{Red nodes constitute a gap.}
  %\label{fig:test1}
\end{minipage}%
\begin{minipage}{.5\textwidth}
  \centering
  \includegraphics[width=.95\linewidth]{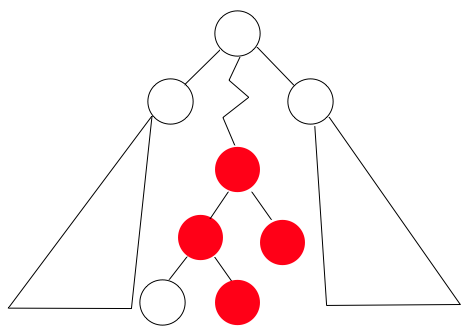}
  %\captionof{figure}{Red nodes constitute a gap.}
  %\label{fig:test2}
\end{minipage}
\caption{The subtree with red nodes on the left is a gap in the complete subtree model, since it is a complete subtree. The subtree with red nodes on the right is not a gap in the complete subtree model. However, it is a gap in the general model.}
\label{fig:complete subtree model vs general model}
\end{figure}

Every gap in the complete subtree model is a gap in the general model, but not vice versa (see Figure \ref{fig:complete subtree model vs general model}). 

Touzet \cite{Touzet2003} computed the complete subtree gap edit distance using a product tree data structure. We present a different algorithm that is motivated by sequence alignment with gaps (\ref{thm:sequence alignment with gaps}) and classic edit distance of Zhang and Shasha \cite{ZhangShasha1989}. In particular, we prove that:

\begin{theorem}[Complete Subtree Gap Tree Edit Distance]
\label{thm:complete subtree gap edit distance with affine gap penalty function}
Given two ordered labeled binary trees $T_{1}$ and $T_{2}$ with vertex set $V_{1}: = V(T_{1})$ and $V_{2}: = V(T_{2})$ respectively, and an affine gap cost function. Let $m := |V_{1}|$ and $n := |V_{2}|$. The complete subtree gap edit distance between $T_{1}$ and $T_{2}$ can be computed in $O(m^{2}n^{2})$ time. If $m \geq n$, then the running time is $O(m^{4})$. 
\end{theorem}

Let $m: = |T_{1}|$ and $n:= |T_{2}|$. Order the nodes in $T_{1}$ and $T_{2}$ via preorder traversal for the same reason as in the general gap case: a gap node is continuing a gap if and only if its parent is a gap node. Enumerate the nodes in $T_{1}$ as $1, 2, \cdots, m$, and the nodes in $T_{2}$ as $1, 2, \cdots, n$, and identify each node (together with the labeling) with its index in this preorder ordering. Let $T_{1}[i'..i]$, $T_{2}[j'..j]$, $\forestdist(T_{1}[i'..i], T_{2}[j'..j])$, $Q, Q_{\perp *}$ and $Q_{* \perp}$ be the same as in the general gap case. Our goal is again to compute:
\[\gamma(T_{1}, T_{2}) = \gamma(M) = Q[1..m, 1..n].\]

\begin{theorem}[Recurrence of Auxiliary Matrices in Complete Subtree Gap Model]
\label{thm:recurrence for aux functions in complete subtree model}
Given the preorder ordering on the nodes of two ordered labeled trees $T_{1}$ and $T_{2}$. Fix nodes $i_{1}\in V_{1}, j_{1}\in V_{2}$. For any $i\in \desc(i_{1})$ and $j\in \desc(j_{1})$, we have the following recurrence relations:
\begin{equation}\label{equ:recurrence for Q complete subtree model}
Q[i_{1}..i, j_{1}..j] = \min 
\begin{cases}
Q[i_{1}..i-1, j_{1}..j-1] + p(i, j)\\
Q_{\perp *}[l_{1}..i, j_{1}..j]\\
Q_{*\perp}[i_{1}..i, j_{1}..j] 
\end{cases}
\end{equation}
%\vspace{-7ex}
\begin{equation}\label{equ:recurrence for Qperpstar complete subtree model}
Q_{\perp *}[i_{1}..i, j_{1}..j] = \min
\begin{cases}
Q[i_{1}..i-1, j_{1}..j] + (a + b) \\
Q_{\perp *}[i_{1}..p(i), j_{1}..j] + b(i - p(i))
\end{cases}
\end{equation}
%\vspace{-5ex}
\begin{equation}\label{equ:recurrence for Qstarperp complete subtree model}
Q_{*\perp}[i_{1}..i, j_{1}..j] = \min
\begin{cases}
Q[i_{1}..i, j_{1}..j-1] + (a + b)\\
Q_{*\perp}[i_{1}..i, j_{1}..p(j)] + b(j - p(j))
\end{cases}
\end{equation}
Here $p(i)$ (resp. $p(j)$) is the index of parent node (if exists) of $i$  (resp. $j$).\end{theorem}

\begin{proof}
We prove recurrence (\ref{equ:recurrence for Q complete subtree model}) and (\ref{equ:recurrence for Qperpstar complete subtree model}). Recurrence (\ref{equ:recurrence for Qstarperp complete subtree model}) can be obtained by a symmetric argument.

$\bullet$ In the first recurrence of $Q[i_{1}..i, j_{1}..j]$, there are three cases:

(1) None of $i$ or $j$ is a gap point, then $i$ must be matched to $j$ (\verb+prove this+!) and thus 
\[Q[i_{1}..i, j_{1}..j] = Q[i_{1}..i-1, j_{1}..j-1]+ p(i,j),\] 
where $p(i, j)$ is the cost of matching $i$ with $j$.

(2) $i$ is a gap point, then 
\[Q[i_{1}..i, j_{1}..j] = Q_{\perp *}[i_{1}..i, j_{1}..j].\]

(3) Similarly if $j$ is a gap node, then 
\[Q[i_{1}..i, j_{1}..j] = Q_{*\perp}[i_{1}..i, j_{1}..j].\]

The above exhaust all the possibilities, hence proves (\ref{equ:recurrence for Q complete subtree model}).

$\bullet$ Next we prove the second recurrence in which $i$ is a gap node. Let $p(i)$ be the parent of $i$. If $i$ is the root then $p(i): = \emptyset$. There are two cases:

(1) If $p(i) = \emptyset$ or $p(i)$ exists and is not a gap node, then $i$ is starting a new gap and hence gets penalized with $a + b$: 
\[Q_{\perp *}[i_{1}..i, j_{1}..j] = Q[i_{1}..i-1, j_{1}..j] + (a + b).\]

(2) If $p(i)$ exists and is a gap node (see Figure \ref{fig:complete_subtree_gap_recurrence}). Then by the complete subtree gap model, every descendent of $p(i)$ must be a gap node as well. There are $i - p(i)$ such nodes. Since they are all continuing a preexisting gap, the total cost is $b(i - p(i))$. In particular if $i$ is the only child of $p(i)$, then $i - p(i) = 1$: we only penalize the node $i$.

\begin{figure}[!htb]
\centering
\includegraphics[scale = .45]{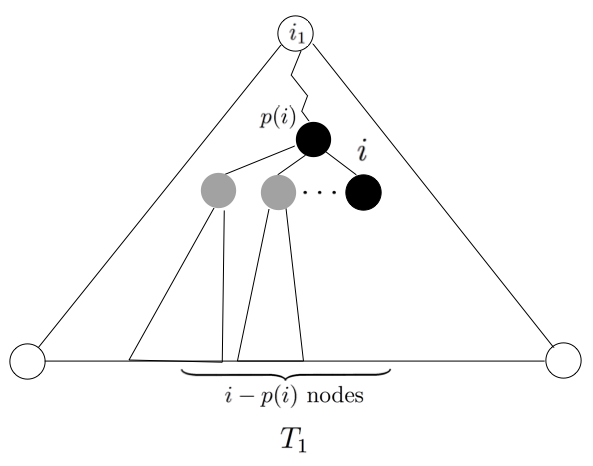}
\caption{$p(i)$ and $i$ are both gap nodes. All the nodes in between them are gap nodes as well.}
\label{fig:complete_subtree_gap_recurrence}
\end{figure}

In this case, we have:
\[Q_{\perp *}[i_{1}..i, j_{1}..j] = Q_{\perp *}[i_{1}..p(i), j_{1}..j] + b(i - p(i)).\]
\end{proof}

\emph{Algorithm for computing $Q[1..m, 1..n]$}:
Let $\treedist(i, j): = Q[i..r(i), j..r(j)]$. Thus, the edit distance between $T_{1}$ and $T_{2}$ in the general gap model is $\treedist(m, n)$, and can be computed by Algorithm \ref{alg:tree edit distance compete subtree gap model dynamic algorithm} below.
\begin{algorithm}
\caption{Complete Subtree Gap Tree Edit Distance}\label{alg:tree edit distance compete subtree gap model dynamic algorithm}
\label{Fingerprint}
\begin{algorithmic}
\State Input: Tree $T_1$ and $T_2$
\State Output: $\treedist(i_{1}, j_{1})$, where $1 \leq i_{1} \leq m: = |T_1|$, and $1 \leq j_{1} \leq n: = |T_2|$
\State Preprocessing: Compute the index of the parent of each node and the $r$ function
\For {($i_{1}, j_{1} = 1$; $i_{1}\leq m, j_{1}\leq n$; $i_{1}$++, $j_{1}$++)}
    	\For{($i = i_{1}$; $i \leq r(i_{1})$;  $i$++)}
		\For{($j = j_{1}$; $j \leq r(j_{1})$;  $j$++)}
        \State Compute $\treedist(i_{1}, j_{1})$ by first compute $Q_{\perp *}[i_{1}..i, j_{1}..j]$,\\ \quad \quad \quad \quad \, then compute $Q_{* \perp}[i_{1}..i, j_{1}..j]$
        		\EndFor
	\EndFor
\EndFor
\end{algorithmic}
\end{algorithm}

It is easy to see that the running time of this algorithm is $O(m^{2}n^{2})$, which proves Theorem \ref{thm:complete subtree gap edit distance with affine gap penalty function}.

%%%%%%%
\section{Application of Complete Subtree Gap Tree Edit Distance to Terrain Comparisons}
\label{sec:application to contour tree comparisons}

\begin{figure}
\centering
\begin{minipage}{.5\textwidth}
  \centering
  \includegraphics[width=.9\linewidth]{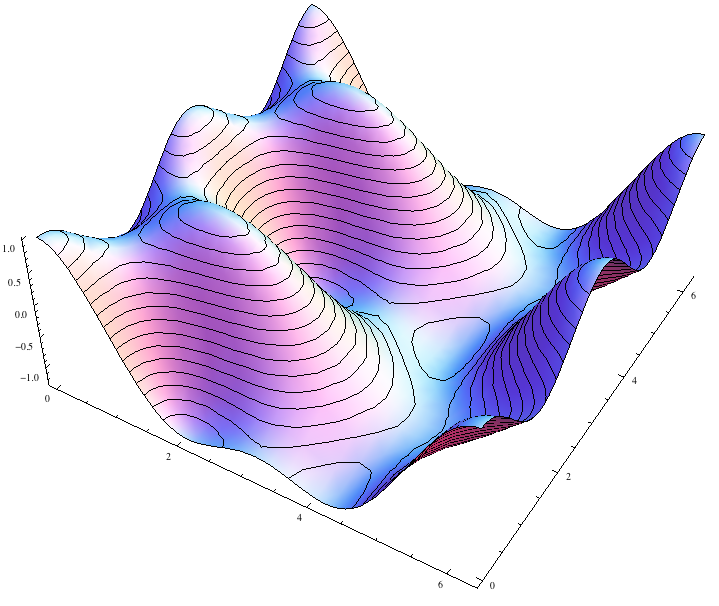}
  \caption{(a)}
\end{minipage}%
\begin{minipage}{.5\textwidth}
  \centering
  \includegraphics[width=.95\linewidth]{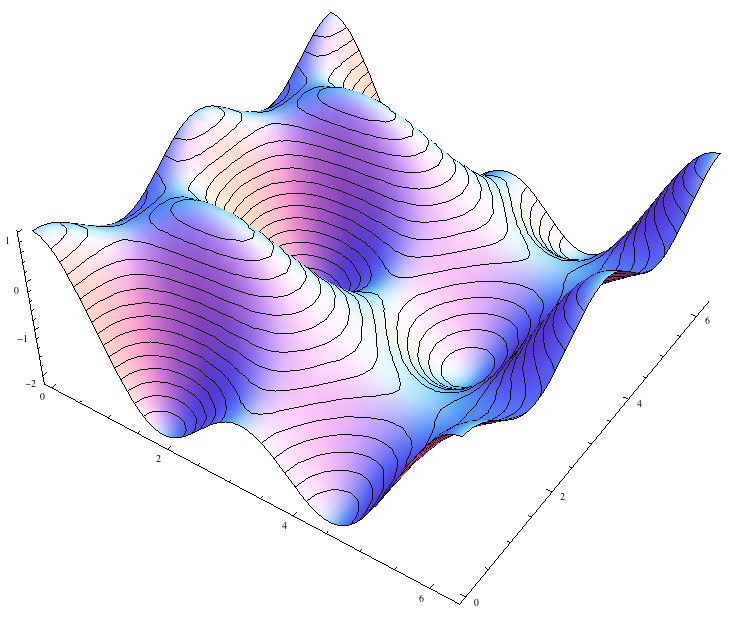}
  \caption{(b)}
\end{minipage}
\caption{Two smooth terrains that look similar. Both of them are local graphs of combinations of trigonometry functions.}
\label{fig:comparing two terrains}
\end{figure}

In this section, we study the problem of comparing the similarities between two terrains. As discussed in the introduction (Section \ref{subsec:persistence and contour tree}), contour trees are the underlying tree structures of terrains that capture the evolution of the connected components of the level sets, or contours. Thus the problem of comparing two terrains (see Figure \ref{fig:comparing two terrains}) can be reduced to comparing their corresponding contour trees.

Roughly speaking, a terrain in $\RR^{3}$ is the graph of a height function on $\RR^{2}$. More precisely, given any smooth Morse function $f: \RR^{2}\longrightarrow \RR^{\geq 0}$ with isolated critical points. The graph of $f$, denoted as $\Gamma_{f}$, is called a smooth terrain. 

Define $M_{<h}(f): = \{x\in \RR^{2} | f(x) < h\}$ to be the points in the plane with height less than $h$, and $M_{h}(f): = \p M_{<h}(f) = \{x\in \RR^{2} | f(x) = h\} $ to be the $h$-level set of $f$. A connected component of $M_{h}(f)$ is called a \emph{contour}. 

\begin{figure}[!htb]
\centering
\includegraphics[scale = .37]{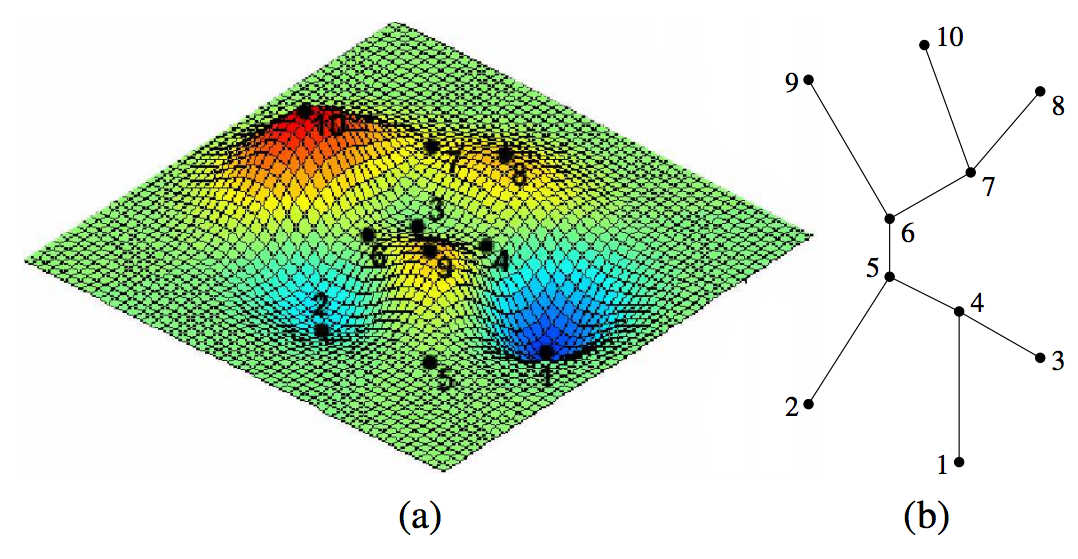}
\caption{Smooth terrain $(a)$ with its contour tree $(b)$. Picture courtesy of  P. Agarwal, L. Arge and K, Yi \cite{Agarwaletal2006}.}
\label{fig:terrain with its contour tree}
\end{figure}

As we vary $h$, the $h$-level set changes topology only at critical points of $f$, which are local maximum, local minimum and saddle points. A contour first appears at a local minimum, disappears at a local maximum. At a saddle point, either two contours join and become a single contour, in which case the saddle is called \emph{negative}; or one contour splits into two, in which case the saddle is called \emph{positive}.
The contour tree of a terrain is defined as follows:

\begin{mydef}[Contour Tree]\label{def:contour tree}
Given a smooth terrain $\Gamma_{f}$ defined as above. The associated contour tree is a graph $C_{f}$ whose nodes are critical points of $f$, and there is an edge $(u, v)$ if a contour appears at $v$ and disappears at $u$. See Figure \ref{fig:terrain with its contour tree}.
\end{mydef}

The contour tree of a terrain was first defined by Boyell and Ruston \cite{BoyellRuston1963}, and is in fact a tree (see \cite{Carretal2003}). Many research has been done \cite{Kreveldetal1997, TarasovVyalyi1998, Agarwaletal2006} in computing the contour tree of a terrain in $\RR^{3}$, and in higher dimensions \cite{CarrSnoeyinkAxen2000, Carretal2003}. Applications of the contour trees have been studied by \cite{GoldCormack1986, KweonKanade1994, SirarCebrian1986, Takahashietal1995}. However, to our best knowledge, there have been no study on the problem of comparing contour trees as a similarity measure of their corresponding terrains. 

In the following, we only consider \emph{piecewise linear} terrains, i.e. graphs of piecewise linear height functions. More precisely, let $\MM$ be a triangulation of $\RR^{2}$, and let $V$ be the set of all vertices in $\MM$. Consider a height function
\[f: V\longrightarrow \RR^{\geq 0}\]
defined on the vertex set, such that $f$ is one-to-one (i.e. no two vertices have the same height). Extend $f$ to the entire plane in a piecewise linear fashion, and identify $f$ with its extension. Thus $\Gamma_{f}$ a is piecewise linear terrain. In this case, all contours are closed polygonal curves. 

Given two triangulations $M_{1}$ and $M_{2}$ of $\RR^{2}$, and height functions $f$ and $g$ defined on the vertices and then extended linearly to $\RR^{2}$. We define the distance between $\Gamma_{f}$ and $\Gamma_{g}$ to be the edit distance between the contour trees $C_{f}$ and $C_{g}$.

Recall that gaps are introduced in tree edit distance to be able to deal with noise in the input. Now the questions is: What should the gap model be in the case of contour trees? It's easy to see that noise in the input terrain (i.e. ``wiggles'' on the surface) are reflected as complete subtrees in the contour tree. 

Here is the upshot: complete subtree gap edit distance can be used to compute the similarities between two contour trees, which can then be used as a measure of the original terrains. It is worth noting that this is only a topological measure of similarity, since the contour tree is a topological construct. Two terrains with identical contour trees do not need to share the same geometry (e.g. curvature, area, geodesics, etc).

There are several natural candidates for the cost of gaps. A topological penalty could be the persistence of noise in the terrain that corresponds to a gap in the contour tree. Geometric penalties could be the height or the volume of the noise in the terrain that correspond to a gap. We leave the understanding of which penalty function is better as well as the implementations to a future project. 
%%%%%%%

\section{Further Improvements}
\label{sec:further improvements}

In both the general and the complete subtree gap models, gaps can have arbitrary sizes (up to the size of the tree). However in some applications, one usually has an upper bound on the size of the noise in the input, and hence on the size of gaps. A natural generalization is to incorporate this upper bound criteria in these gap models. Given a tree $T$ and an integer $k$ such that $0 < k \leq |T|$. Define a gap $g$ to be an arbitrary subtree with at most $k$ nodes. When $k = |T|$, this is the general gap model. Consequently in our recurrences, when a node is continuing a gap, we need the additional check on whether the current gap size has exceeded $k$ or not before penalizing the gap node. We leave more rigorous discussions on this improvement to a future project.
\chapter{Conclusion and Future Works}
\label{chap:conclusion and future works}

In this thesis, we studied edit distance with gaps between two ordered, labeled trees. Touzet \cite{Touzet2003} proposed two gap models: the general model and the complete subtree model. Given two trees $T_{1}$ and $T_{2}$ with $m$ and $n$ nodes respectively. We computed the general gap edit distance between binary trees in $O(m^{3}n^{2} + m^{2}n^{3})$ time, and the complete subtree gap edit distance between arbitrary trees in $O(m^{2}n^{2})$ time. Our dynamic programming algorithms are motivated by the classic sequence alignment \cite{SetubalMeidanis1997} algorithms and Zhang and Shasha's classic edit distance algorithm \cite{ZhangShasha1989}. In both models, we assume that the gap cost function is affine. Prior to our work, no explicit algorithm was known in computing the general gap edit distance, since such computation is NP-hard for arbitrary trees (see \cite{Touzet2003}). We studied an application of the complete subtree gap edit distance in terrain comparison via comparing the similarities between the corresponding contour trees.

The following are some open problems that are suitable for a future project:
\begin{problem}
Recently S. Sankararaman, P. Agarwal and T. Molhave \cite{Agarwal2013} studied the problem of comparing similarity between two trajectories sampled at a certain rate, using sequence alignment, which is a topological construction, and dynamic time warping, which is a geometric construction.

The gapped edit distance is by definition a topological measure of similarities between trees. Is is possible to combine the edit distance with some geometric similarity measures (e.g. dynamic time warping) as in the trajectory alignment case?
\end{problem}

\begin{problem}
Our algorithm for computing the general gap edit distance between binary trees seems to suggest that the NP-hardness of computing this distance for arbitrary trees comes from the fact that the degrees or the branching factors of the internal nodes vary. Moreover, the running time should depend on the degree in an exponential fashion. A natural next step is to compute this distance for trees with fixed degrees (e.g. ternary trees).
\end{problem}

\begin{problem}
Our algorithms are too slow for any practical applications. Is it possible to simplify the algorithm by recognizing repetitions in the recurrences as in Zhang and Shasha's work \cite{ZhangShasha1989}? 
\end{problem}
%\include{{./Chapter5/chLorem2}}
%==============================================================================

%-----------------------------------------------------------------------------%
% APPENDICES -- OPTIONAL. These are just chapters enumerated by Appendix A,
%                Appendix B, Appendix C...
%-----------------------------------------------------------------------------%
%\appendix
%\include{{./Appendix1/chLorem3}} % Start with '\chapter{Title}'
%You can always add more appendices here if you want

%-----------------------------------------------------------------------------%
% BIBLIOGRAPHY -- uncomment \nocite{*} to include items in 'mybib.bib' file
% that aren't cited in the text.  Change the style to match your
% discipline's standards.  Of course, if your bibliography file isn't called
% 'mybib.bib' you might want to change that here too :)
%-----------------------------------------------------------------------------%
\nocite{*} 
% - if you use this it will put EVERYTHING in your .bib file into the references even if you don't cite it in the text
%\bibliographystyle{./Bibliography/jasa} %Formats bibliography
\bibliographystyle{plain}
\cleardoublepage
\normalbaselines %Fixes spacing of bibliography
\addcontentsline{toc}{chapter}{Bibliography} %adds Bibliography to your table of contents
\bibliography{./Bibliography/mybib} %your bibliography file - change the path if needed
%-----------------------------------------------------------------------------%

%-----------------------------------------------------------------------------%
% BIOGRAPHY -- Start file with '\biography'.  Mandatory for Ph.D.
%-----------------------------------------------------------------------------%
\biography

Hangjun Xu was born in August 1987 in Hangzhou, China. In 2005, He went to Zhejiang Unviersity in China and obtained a Bachelor of Science degree in Mathematics and Applied Mathematics, and the certificate of the \emph{Chu Kochen Honors Program} in 2009. After that he got the graduate student fellowship from Duke University to pursue a Ph.D. degree in Mathematics since 2009, and his field of research has been differential geometry and geometric algorithms. He started pursuing a Master's degree in Computer Science en route to his Ph.D. program since Fall, 2011. In 2012, he won the Graduate School Research Fellowship for $5000$ dollars. Since Fall 2010, he has taught $8$ undergraduate courses as an instructor, including calculus I and II, linear algebra, ordinary and partial differential equations. After graduation, he will be working as a senior software engineer at Oracle in Santa Clara, California.

%-----------------------------------------------------------------------------
% You're done :)
\end{document}